\documentclass[aip,reprint]{revtex4-1}

\usepackage{bm}
\usepackage[mathlines]{lineno}

\usepackage[utf8]{inputenc}
\usepackage[T1]{fontenc}
\usepackage{mathptmx}

\usepackage{amsmath}
\usepackage{xfrac}
\usepackage[inline]{enumitem}   
\usepackage{gensymb}
\usepackage{amssymb}
\usepackage{rotating}
\usepackage{pgfplots}
\usepackage{relsize}
\usepackage{hyperref}
\usepackage{pgfplotstable}
\usetikzlibrary{patterns}
\usetikzlibrary{arrows.meta,positioning,pgfplots.groupplots,calc,angles,positioning,intersections,quotes,math,matrix,spy}
\usepgfplotslibrary{statistics}

\draft 

\begin{document}
	
	
	
	\title{Decreasing the uncertainty of classical laser flash analysis using numerical algorithms robust to noise and systematic errors}
	
	
	
	\author{Artem Lunev}
	\email[Electronic address: ]{artem.lunev@ukaea.uk}
	\homepage[Download PULsE from: ]{https://kotik-coder.github.io/}
	\affiliation{United Kingdom Atomic Energy Authority~(UKAEA), Culham Science Centre, Abingdon, Oxfordshire OX14 3DB, United Kingdom}
	
	\author{Robert Heymer}
	\affiliation{United Kingdom Atomic Energy Authority~(UKAEA), Culham Science Centre, Abingdon, Oxfordshire OX14 3DB, United Kingdom}
	
	\date{\today}
	
	\begin{abstract}
		The laser flash method is highly regarded due to its applicability to a wide temperature range, from cryogenic temperatures to the melting point of refractory metals, and to extreme environments involving radioactive or hazardous materials. Although instruments implementing this method are mostly produced on a commercial basis by major manufacturers, there is always room for improvement both in terms of experimental methods and data treatment procedures. The measurement noise, either due to the detector performance or electromagnetic interferences, presents a significant problem when accurate determination of thermal properties is desired. Noise resilience of the laser flash method is rarely mentioned in published literature; there are currently no data treatment procedures which could guarantee adequate performance under any operating conditions. In this paper, a computational framework combining finite-difference solutions of the heat conduction problem with nonlinear optimization techniques based on the use of quasi-Newton direction search and stochastic linear search with the Wolfe conditions is presented. The application of this framework to data with varying level of noise is considered. Finally, cross-verification and validation using an external standard, a commercial and an in-house built laser flash instrument are presented. The open-source software implementing the described computational method is benchmarked against its industrial counterpart.   
	\end{abstract}
	
	\pacs{}
	
	\maketitle 
	
	\section{Introduction} \label{sect:intro}
	
	Nearly sixty years have passed since~\citet{Parker1961} first proposed the flash method for determination of thermal properties. Unlike other methods, e.g. the transient hot-strip method~\cite{Gustafsson_1979}, the transient hot wire method~\cite{Hammerschmidt2000}, and the recently revisited guarded hot plate method~\cite{Salmon_2001,Jannot_2010,Scoarnec2015}, which suffer from an inherent thermal contact resistance between the heater and the sample, the flash method implements a scheme for contactless heating by a pulsed laser source. The latter induces a time-dependent temperature response measured typically at the rear-surface of a cylindrical sample~(although other detection concepts have been considered by~\citet{PAVLOV201739} and previously by~\citet{Ronchi1999}) either by an infrared detector or by a thermocouple~(currently, the use of immersive thermocouples is discouraged due to the extra contact resistance introduced in the weld region~\cite{astm2013standard}) and used to infer the thermal properties from a mathematical model of the experiment. Due to its many advantages, such as requiring only small samples, reducing measurement time, and being extendable to very high temperatures~(above~$3000$~K~\cite{Ronchi1999,pavlov2018high,Hay2006}), the flash method soon became a standard in its field. It is truly indispensable for studies of radioactive or hazardous materials~(e.g., low-active proton-irradiated tungsten samples~\cite{HABAINY2018152} and mildly-radioactive neutron-irradiated beryllides~\cite{UCHIDA2003499}), especially for a highly-radioactive material that requires remote access~(\citet{WALKER200619} on the thermal conductivity of a $100$~MWd/kgHM spent oxide fuel from a commercial PWR nuclear reactor), due to the versatility of sample mounting required for a pre-programmed robotic arm, for the use of a manipulator or, more commonly, for a glovebox environment.
	
	The ASTM standard E1461 -- 13~\cite{astm2013standard} sets out the applicability of the laser flash method to `\textit{essentially fully dense (preferably, but low porosity would be acceptable), homogeneous, and isotropic solid materials that are opaque to the applied energy pulse}'. The measurement procedure corresponding to these conditions is henceforth referred to as the `\textit{classical laser flash analysis}' to distinguish it from more sophisticated cases, such as when analyzing liquids and melts~\cite{blumm2003characterization}, semi-transparent and diathermic samples, etc. Assuming ideal operation of the instrument and correct handling of the samples, a simple heat transfer model accounting for the true shape of the laser pulse and the radiative losses at the sample boundaries should perfectly describe raw data. In some cases, the experimental uncertainties add up to the systematic and random errors in input time-temperature profiles to such an extent that the conventional algorithms produce systematically biased or randomly scattered output values. These uncertainties include~(but are not limited to) those of the `\textit{detector performance and of the data acquisition system}', which should deliver a `\textit{linear electrical output proportional to the temperature rise}'. 
	
	Assuming perfect delivery of the laser pulse to the sample's front surface, the performance of a laser flash instrument is influenced by the following factors:
	
	\begin{enumerate}[label=({\roman*})]
		\item The theoretical detector performance is given by its specific detectivity, a parameter equal to the output current~(or voltage) per watt of incident thermal power per unit area of the active surface. The detectivity may vary over several orders of magnitude depending on the detector material and the spectral characteristics of the incident thermal radiation. For instance, when comparing near-infrared Peltier-cooled detectors, photovoltaic MCT~(mercury cadmium telluride) are generally much better in terms of detectivity than the photoconductive PbSe detectors; however, Peltier-cooled detectors perform worse in general than the liquid-nitrogen cooled InSb detectors. When doing measurements at cryogenic or very high temperatures, even the best near-infrared detectors yield inherently noisy data, and detectors suitable for a different spectral range are desired;
		\item The analog signal generated by the detector is processed with a (cooled) pre-amplifier unit mounted within the detector assembly. The preamplified signal is then transmitted to the main amplifier in the processing unit, and later converted to a digital output with the digital-to-analog convertor. Additional electromagnetic noise may arise from the technological process involved in making the electrical connections~(commonly soldering);  
		\item As pointed out by~\citet{Baba_2001}, the synchronization of the laser pulse with the start of data acquisition by the detector may be challenging. This is generally accomplished with the use of a pulse monitor, and if faulty the latter may yield an erroneous systematic shift in the time sequence; 
		\item A poorly transmitting optical window, e.g. separating the detector assembly from the furnace interior, may lead to a drop in the signal quality. This sometimes happens due to the re-deposition of the material released from the sample, e.g. due to a decomposition process at high temperatures;
	\end{enumerate}

	Considering the systematic nature of most occurrences above, it would be impossible to correct them using repeat measurements. Some other uncertainties are intermittent: \citet{Sheindlin1998} mention baseline drift as one possibility; \citet{sramkova1995using} considered oscillations in the mains current~(`hum') and a shift in the baseline when performing a Levenberg-Marquardt minimization procedure. Synthetic high-noise time-temperature profiles were considered by~\citet{carr2019a} and~\citet{CARR2019118609} to show that the standard half-rise time procedure is not always applicable under otherwise ideal experimental condition. 
	
	To the best of authors' knowledge, the listed experimental uncertainties have not been studied with regard to their effect on the output of classical laser flash analysis. Likewise, neither does the ASTM standard set any acceptance criteria for the quality of the instrumental time-temperature profiles, which prevents effective quality control for commercially built instruments. Depending on the design solutions chosen by the manufacturer, the output signal may be of lower quality than required for reliable operation of the conventional data treatment procedures. The goal of this paper is to present a universal and easily verifiable method of data analysis that would:~\begin{enumerate*}[label=(\alph*)]
		\item not use cumbersome analytic expressions;
		\item be independent of general-purpose commercially distributed mathematical packages and software;
		\item be highly resilient to the experimental uncertainties of the classical laser flash analysis
	\end{enumerate*}. This method would then be used to process data from a laser flash experiment in a commercial instrument using an external standard. Ultimately, the results would be benchmarked with the industrial software to asses its capabilities. The algorithm and procedures outlined in this work are part of the PULsE~(\textbf{P}rocessing \textbf{U}nit for \textbf{L}aser Fla\textbf{s}h \textbf{E}xperiments) software, which is an open-source, cross-platform Java code freely distributed under the Apache 2.0 license~\cite{artem_lunev_2019_3482937}. 
	
	\section{General remarks on experimental data treatment}
 
	The computational framework described in this paper implements finite-difference schemes for solving a heat transfer problem, rather than relying on existing semi-analytical solutions by previous authors. Some of the remarks below may be specific to the way this solution is obtained.
	
	\subsection{Calculation of the objective function}
	\label{sect:target_function}
	
	In the reverse heat conduction problem, two enumerated collections are compared against each other: the experimental data sequence~$\Delta T(\widetilde{t}_i)$, $i = 0,1,...,n_{\rm{exp}}$, where~$n_{\rm{exp}} \simeq 1,000-5,000$ is the number of experimental data points; and the calculated dataset~(containing only unique elements) representing the heating curve~$\Delta \widehat{T}(t_j)$, $j = 0,1,...,n_{\rm{s}}$, where an arbitrary~$n_{\rm{s}} \simeq 100$ is chosen. Here we assume fixed time sampling for the latter, so that $\exists \ \Delta t_{\rm{s}} \in \mathbb{R}, \ \forall \ j=0,...,n_{\textrm{s}}-1: \ t_{j+1} - j_i = \Delta t_{\rm{s}}$. Note that generally $\widetilde{t}_i \neq t_j$, i.e. the time values for the experimental and calculated data points may not overlap.
	
\begin{figure}[!ht]
	\centering
	\begin{tikzpicture}
	\pgfplotsset{
		xmin=25.25, xmax=28.10,
		ymin=6.25, ymax=6.6,
		xlabel={Time, $t$},
		xlabel near ticks,
		ylabel near ticks,
		ylabel={Temperature rise, $\Delta T$},
		xtick=\empty,
		ytick=\empty,
		axis lines=left,
		width=1.0\columnwidth,
		legend cell align={left}
	}
	\begin{axis}[domain=25.5:26.0,restrict y to domain=6.1:6.8,legend pos=south east]
	\addplot[very thick,color=red,mark=square*,only marks,mark options={scale=1.0}] 
	table[x expr=\thisrowno{0}*1000,y expr=\thisrowno{1} + 8.51379655] {ExperimentalData_01}
	node[pos=0.074,pin=left:$\Delta T(\widetilde{t}_{i})$]{};				
	\addlegendentry{Experimental data};
	\addplot[color=green,mark=otimes*, mark options={scale=1.5},every node near coord/.style={anchor=90}] 
	table[x expr=\thisrowno{0}*1000,y=Temperature] {HeatingCurve_01}
	node[pos=0.1945,pin=below:$\Delta \widehat{T}(t_{k})$]{}
	node[pos=0.302,pin=below:$\Delta \widehat{T}(t_{k+1})$]{}
	node[pos=0.3988,pin=below:$\Delta \widehat{T}(t_{k+2})$]{};
	\addlegendentry{Model solution};
	\addplot[color=olive,mark=*, mark options={scale=1.5},only marks,fill=white] 
	table[x expr=\thisrowno{0}*1000,y expr=\thisrowno{1}] {InterpolatedData_01}		
	node[pos=0.21,pin=left:$\Delta \widehat{T}(\widetilde{t}_{i})$]{};				
	\addlegendentry{Interpolated points};
	\addplot[color=black,mark=\empty, dashed] 
	table[x expr=\thisrowno{0}*1000,y expr=\thisrowno{1}] {Difference};
	\end{axis}		
	\end{tikzpicture}
	\caption{\label{fig:interpolation}An illustration of calculating the sum of squared residuals~(SSR) using data from two enumerated collections of different size: the experimental data points~$\Delta T ( \widetilde{t}_i )$~($i=0,1,...,n_{\rm{exp}}$) and the model solution~$\Delta \widehat{T} ( t_j )$~($j=0,1,...,n_{\rm{s}}$). Eq.~\ref{eq:interpolated} is used to calculate the interpolated value~$\Delta \widehat{T} ( \widetilde{t}_i )$ at~$t_k < \widetilde{t} < t_{k+1}$ using the nearest points of the model solution~$\Delta \widehat{T} ( {t}_{k} )$ and $\Delta \widehat{T} ( {t}_{k+1} )$ and the experimental time~$\widetilde{t}_i$.}
	\end{figure}
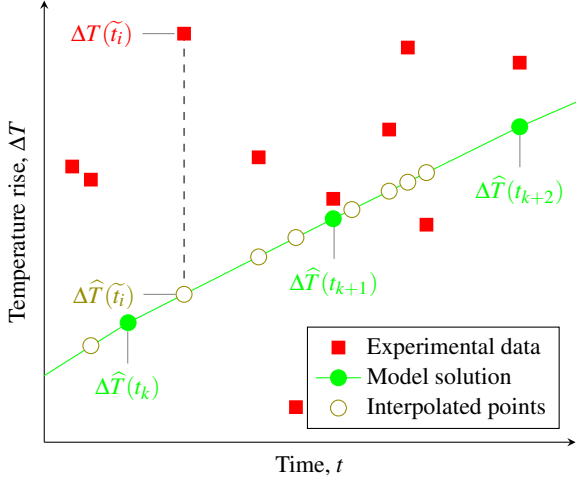

	To define a computational algorithm for the objective function, it is thus necessary to implement an interpolation procedure. Instead of interpolating over the \textit{experimental} dataset~$\Delta T(\widetilde{t}_i)$, which is guaranteed to produce interpolation errors~(especially in the case of noisy detector data), the idea is to interpolate over the \textit{solution} values~$\Delta \widehat{T}(t_j)$~(Fig.~\ref{fig:interpolation}). With linear interpolation using nearest adjacent points, the value of the solution at time~$\widetilde{t}_i$, $i = 1,...,n_{\rm{s}}-1$ can be calculated as:
	\begin{eqnarray}
	\label{eq:interpolated}
	\Delta \widehat{T}(\widetilde{t}_i) = 
	\frac{t_{k+1}-\widetilde{t}_i}{\Delta t_{\rm{s}}} \times \Delta \widehat{T}(t_k) + \nonumber \\ \frac{\widetilde{t}_i-t_{k}}{\Delta t_{\rm{s}}} \times \Delta \widehat{T}(t_{k+1}) , \ k =  \left \lfloor \frac{\widetilde{t}_i}{\Delta t_{\rm{s}}} \right \rfloor,
	\end{eqnarray}
	where square brackets denote the floor function, i.e. the greatest integer less than or equal to the argument, and~$k \in [0,n_{\rm{s}}] \cap \mathbb{Z}$ is the index of the greatest element~${t}_k$ in the calculated dataset such that~$t_k \leq \widetilde{t_i}$.
	
	The objective function~$f$ can then be calculated as the sum of squared residuals~(SSR) as follows:
	\begin{equation}
	\label{eq:ssr}
	f = \sum_{i=a}^{b} \left( \Delta T(\widetilde{t}_i) - \Delta \widehat{T}(\widetilde{t}_i) \right)^2,
	\end{equation}
	where $a,b \in [0,n_{\rm{exp}}] \cap \mathbb{Z}$ are the lower and higher sum limits, respectively, defining the time domain for calculation.
	
	\subsection{Conversion of the detector voltage to relative heating}
	\label{sect:conversion}
	
	The voltage~$U(t)$ transmitted from the infrared detector to the data acquisition module determines the spectral radiance~[$\rm{W} \times \rm{sr}^{-1} \times m^{-3}$], which is used to estimate the temperature rise of the sample's rear surface. Alternatively, the voltage $U(t)$ from a thermocouple generated due to the Seebeck effect is converted to temperature via a characteristic function of the thermocouple. Whatever the tool used to measure the temperature,  hereinafter it is referred to as \textit{the detector}, and the voltage it measures is referred to as~\textit{the signal}.
	
\begin{figure*}
	\centering
	\begin{tikzpicture}
	\pgfplotsset{
		xlabel near ticks,
		ylabel near ticks,
		legend cell align={left}
	}
	\begin{axis}[xmin=-80, xmax=95.0,
	ymin=-10.0, ymax=4.0,
	xlabel={Time, $t$ (ms)},
	ylabel={Detector signal (mV)},
	width=0.5\textwidth,
	legend style={at={(rel axis cs: 1.25, 1.0)}, anchor=north west,legend columns=1,fill=none,align=center,thin,},	
	set layers=standard,
	]
	\addplot[very thick,color=red,mark=o,only marks,mark options={scale=0.25}] 
	table[x expr=\thisrowno{0}*1000,y expr=\thisrowno{1}] {ExperimentalData_01};
	\addlegendentry{Experimental data};
	\addplot [color=blue,thick,domain=-100:100, samples=20]{-8.51379655};
	\addlegendentry{Constant baseline};
	\addplot [color=blue,thick,domain=-100:100, samples=20,loosely dotted]{x*6.8086E-3-8.24724};
	\addlegendentry{Linear baseline};
	\addplot [color=green,very thick,domain=-80:95, samples=20,on layer={axis foreground}]{7.05636-8.51379655};
	\addlegendentry{Averaged maximum};
	\coordinate (Ac) at (axis cs:-25.0,-8.51379655);
	\coordinate (Bc) at (axis cs:-25.0,-1.45743655);
	\coordinate (Cc) at (axis cs:-31.0,-5.1743655);
	\node (A) at (Ac) {};
	\node (B) at (Bc) {};
	\node[rotate=90] (C) at (Cc) {$\Delta U_{\rm{max}}$};
	\draw[thick,dashed,{Stealth}-{Stealth}] (A) -- (B);
	\addplot[very thick,color=cyan,mark=\empty,on layer={axis foreground}] 
	table[x expr=\thisrowno{0}*1000,y expr=\thisrowno{1}] {RunningAverage};
	\addlegendentry{Running average};
	\coordinate (POS1) at (rel axis cs:1.25,0.0);	
	\end{axis}		
	\begin{axis}[
	ybar,
	ymin=0,
	at={(POS1),anchor=south west},
	width=0.325\textwidth,
	x,
	ylabel={Probability density},		
	ylabel style={font=\footnotesize},
	xlabel style={font=\footnotesize},
	ticklabel style = {font=\footnotesize},
	]
	\addplot +[
	hist={
		density,
		bins=32,
		data min=-10,
		data max=5,
	}   
	] table [y index=1] {ExperimentalData_01}
	node[pos=0.5275,pin={right:$\overline{U}_{\rm{p}}$}]{}
	node[pos=0.08,pin={above:$\overline{U}_{\rm{min}}$}]{};
	\coordinate (AAc) at (axis cs:7.05636-8.51379655,-0.5);
	\coordinate (BBc) at (axis cs:7.05636-8.51379655,1.0);
	\node (AA) at (AAc) {};
	\node (BB) at (BBc) {};
	\draw[green,densely dashed,very thick] (AA) -- (BB);	
	\end{axis}
	\end{tikzpicture}
	\caption{\label{fig:baseline_scaling}An example pre-processing of the experimental data~(PbSe detector), showing: the baseline calculated by least-squares fitting to experimental data at~$t < 0$; a running average curve generated by coarsening of experimental data using a default reduction factor of~$32$ and thus ensuring relative robustness to outliers; and the maximum~$\overline{U}_{\rm{max}}$ of the running average, which roughly corresponds to the~$\overline{U}_{\rm{p}}$ peak in the probability density function.}
	\end{figure*}
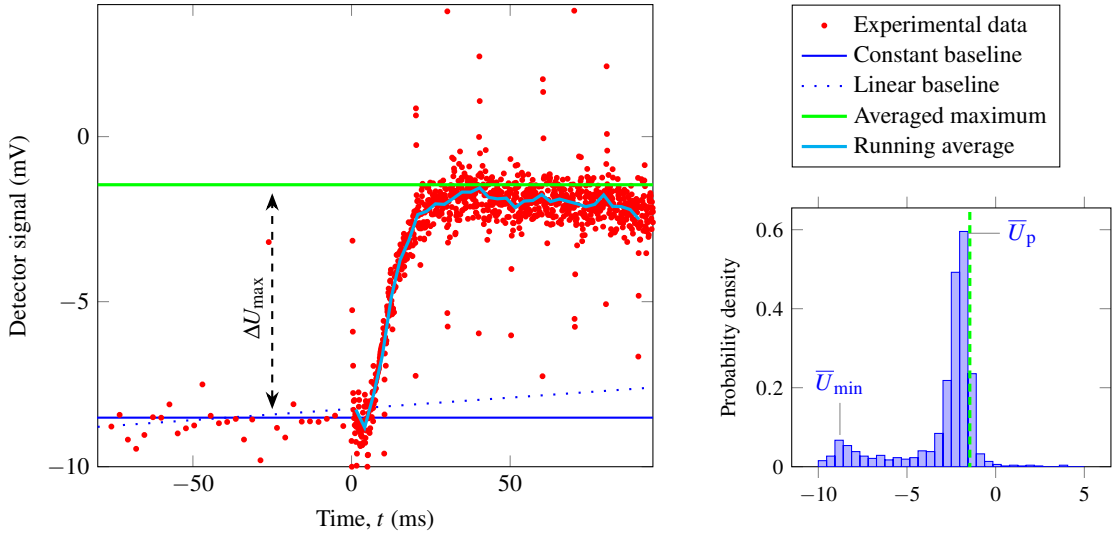

	A linear relation between the signal~$U(t)$ and the heating~$\Delta T(t) = T(t) - T_0$, where $T_0$ is the baseline test temperature measured by a separate detector, is assumed: $\Delta T(t)/C_2 = U(t) - C_1$, where the constant $C_1$ can be determined as the baseline level~$U_{\rm{min}}$ and $C_2$ -- as the maximum relative heating~$\Delta T_{\rm{max}} \propto \Delta U_{\rm{max}}$~(Fig.~\ref{fig:baseline_scaling}). To calculate the absolute temperature in degrees, a further conversion is necessary, e.g. for infrared detectors using Planck's equation and the emissivity~$\varepsilon (T)$ of the sample. When a nonlinear dependence of temperature on the spectral radiance may be neglected, e.g. at high temperatures for infrared detectors~(\citet{Baba_2001}) or when using a thermocouple, thermal diffusivity may be calculated without introducing additional error utilizing just the heating values measured in arbitrary units.
	
	It is important though to correctly estimate the baseline signal~$U_{\rm{b}}(t)$ and the maximum change of voltage~$\Delta U_{\rm{max}}$, assumed to be proportional to the maximum heating~$\Delta T_{\rm{max}}$. This becomes challenging for some experimental assemblies with noisy~(see Fig.~\ref{fig:baseline_scaling}) and/or drifting signal~(e.g. due to unstable detector current or imperfect temperature regulation). A linear baseline~$\Delta T_b(t) = \Delta T_{\rm{lin}} + k_{\rm{lin}} \cdot t$ is usually enough to accommodate the drift, with~$\Delta T_{\rm{lin}}$ and~$k_{\rm{lin}}$ determined using a simple linear regression for the detector signal acquired at~$\widetilde{t}_i < \widetilde{t}_{i0} := 0$, i.e. before the laser pulse: 
	\begin{subequations}
		\begin{align}
		&\Delta T_{\rm{lin}} = \left < T \right > - k_{\rm{lin}} \cdot \left < {t} \right >, \\
		&k_{\rm{lin}} = \mathlarger{\sum}_{i=0}^{i_0 - 1} \frac{\left ( \widetilde{t}_i - \left < t \right > \right ) \cdot \left ( T_{\rm{b}}(\widetilde{t}_i) - \left < T \right > \right )}{\left ( \widetilde{t}_i - \left < t \right > \right )^2} \\
		&\left < T \right > = \frac{1}{i_0}\sum_{i=0}^{i_0 - 1}{T_{\rm{b}}(\widetilde{t}_i)}, \
		\left < {t} \right > = \frac{1}{i_0}\sum_{i=0}^{i_0 - 1}{\widetilde{t}_i}.
		\end{align}
	\end{subequations}
	
	As seen from Fig.~\ref{fig:baseline_scaling}, a constant-baseline approach~($k_{\rm{lin}} = 0$) is sometimes preferred when fitting to a limited data sample with no obvious drift, as the random spread of data may be too large for an accurate estimation of the slope. 
	
	The outliers in Fig.~\ref{fig:baseline_scaling} present measurement artefacts and prevent a simple estimation of the maximum heating~$\Delta T_{\rm{max}}$ based on the absolute maximum value of the measured signal. An outlier-robust procedure based on data binning and coarsening~(i.e., calculation of a running average of large chunks of data) has been implemented. Fig.~\ref{fig:baseline_scaling} shows how effective this procedure is in terms of finding the peak in the probability density of the signal distribution with negligible drift of the measured signal, compared to simply finding the absolute maximum of the signal.
	
	\section{Heat transfer model for the laser flash experiment} 
	\label{sect:model}
	
	The idealized heat transfer model introduced by~\citet{Parker1961} refers to the adiabatic heating of the sample material by an infinitesimal laser pulse. The two latter conditions are rarely observed in practice, and hence more sophisticated heat conduction models had been proposed by~\citet{Cowan1963}, \citet{CapeLehman1963}, \citet{ClarkTaylor1975}, and \citet{THERMITUS19974183} who focused on heat losses~(linearized with respect to the temperature rise due to it being small), while~\citet{LarsonKoyama1967}, \citet{Azumi1981}, and \citet{LECHNER1993341} considered primarily the effect of finite pulse widths. In this section, a dimensionless heat conduction model is formulated and justified based on the previous analysis.
	
	When the laser spot uniformly covers the sample's front surface and the sample diameter is much larger than its thickness~($l \ll d$, but typically~$l/d \approx 0.1 - 0.2$), the heat transfer is effectively one-dimensional. Assuming that heating is small, so that the thermal conductivity does not change across the sample's~$x$ coordinate~($\lambda \neq \lambda(x)$) and~$\Delta T/T_0 \ll 1$, the boundary problem may be linearized. Finally, it becomes possible to convert the heat equation and the boundary conditions to a dimensionless form by introducing the Biot~(Bi) number, which determines the heat transfer resistance at sample's surface, and the Fourier~(Fo) number, which characterizes transient heat conduction, as dimensionless quantities, thus allowing to simplify the solution by reducing the number of variables. 
	
	\subsection{Boundary problem}
	\label{sect:boundary_problem}
	
	A one-dimensional heat conduction problem for the laser flash experiment must include: \begin{enumerate}[label=(\roman*)]
		\item a heat source term $QP(t)/(\pi d^2/4)$, where~$Q$ is the heat current, $P(t) \sim 1/t_{\mathrm{las}}$ is the normalized time distribution of the laser pulse, and~$d$ is the sample diameter; 
		\item the heat sink terms at both the front~$x=0$ and the rear~$x=l$ surfaces due to the radiant heat flux~$q_{12} = S_1 \varepsilon_1(T) \sigma_0 T^4(x,t) - S_2 \varepsilon_2(T) \sigma_0 T_0^4$ from the sample's non-concave surface~$S_1$ to the surrounding surface~$S_2$ of the furnace interior~(assuming the latter is kept at a stable temperature~$T_0$), where~$\varepsilon_1$ and~$\varepsilon_2$ are the emissivities of the sample and the surrounding `shell' of an arbitrary shape.
	\end{enumerate}
	
	In addition, heat flow is assumed to be axial, meaning radial heat fluxes of any kind are neglected. Under certain assumptions~(\citet{avduevskii1975fundamentals}), the heat flux~$q_{12} \approx \varepsilon \sigma_0 \left ( T^4(x,t) - T_0^4 \right ) H_{12}$, where~$H_{12}$ is the mutual radiant surface and~$\varepsilon = \left \{ \varepsilon_1^{-1} + (S_1/S_2) (\varepsilon_2^{-1} - 1) \right \}^{-1}$ is the reduced emissivity. Considering that~$S_2 \gg S_1$, the latter expression reduces to simply~$\varepsilon \simeq \varepsilon_1$, so that the boundary problem is written as:
	\begin{subequations}
		\label{eq:boundary_problem}
		\begin{align}
		\label{eq:he_1d_dim}
		&C_{\mathrm{p}} \rho \frac{{\partial T}}{{\partial t}} = \frac{\partial}{\partial x} \left[ \lambda \frac{{{\partial}T}}{{\partial {x}}} \right],\quad {\kern 1pt} 0 < x < l,\quad t > 0,\\
		\label{eq:bc_0x_dim}
		&{\left. {\lambda \frac{{\partial T}}{{\partial x}}} \right|_{x = 0}} =  - \frac{4Q}{{\pi {d^2}}}P(t) + \varepsilon_1 ({T_0})\sigma_0 \left[ {{T^4}(0,t) - T_0^4} \right],\\
		\label{eq:bc_lx_dim}
		&{\left. {\lambda \frac{{\partial T}}{{\partial ( - x)}}} \right|_{x = l}} = \varepsilon_1 ({T_0})\sigma_0 \left[ {{T^4}(l,t) - T_0^4} \right],\\
		\label{eq:ic_dim}
		&T(0,x) = {T_0},
		\end{align}	
	\end{subequations}
	where~$C_{\mathrm{p}}$ is the specific heat at constant pressure and~$\rho$ is the material density.
	
	The dimensionless quantities are then introduced as follows:
	\begin{subequations}
		\label{eq:variables}
		\begin{align}
		\label{eq:fo}
		&{\rm{Fo}} := a{t}/{{{l^2}}}, \\
		\label{eq:bi}
		&{\rm{Bi}} := {{4\sigma_0 \varepsilon_1 T_0^3l}}/{\lambda }, \\
		\label{eq:theta}
		&\theta  := {{\left(T - {T_0}\right)}}/{{\delta {T_{\rm{m}}}}}, \\
		\label{eq:delta_tm}
		&\delta {T_{\rm{m}}} := {4Q}/{ \left( {C\rho \pi {d^2}l} \right) }, \\
		\label{eq:y}
		&y := {x}/{l}.
		\end{align}
	\end{subequations}
	
	Assuming that~$\lambda \neq \lambda(x)$ and~$(T-T_0)/T_0 \ll 1$, linearization of Eqs.~\eqref{eq:boundary_problem} results in an alternative boundary problem:
	\begin{subequations}
		\label{eq:boundary_problem_dimensionless}
		\begin{align}
		\label{eq:he_1d_nodim}
		&\frac{{\partial \theta }}{{\partial {\rm{Fo}}}} = \frac{{{\partial ^2}\theta }}{{\partial {y^2}}},\quad {\kern 1pt} 0 < y < 1,\quad {\rm{Fo}} > 0,\\
		\label{eq:bc_0x_nodim}
		&{\left. {\frac{{\partial \theta }}{{\partial y}}} \right|_{y = 0}} = {\rm{Bi}} \cdot \theta  - \Phi \left( \rm{Fo} \right),\\
		\label{eq:bc_lx_nodim}
		&{\left. {\frac{{\partial \theta }}{{\partial ( - y)}}} \right|_{y = 1}} = {\rm{Bi}} \cdot \theta,\\
		\label{eq:ic_nodim}
		&\theta (0,y) = 0.
		\end{align}	
	\end{subequations}
	
	\subsection{Laser pulse function}
	\label{sect:pulse_function}
	
	The laser pulse functions can be changed programmatically during the experiment and can take various forms, including the following as example~(a rectangular, a triangular, and a Gaussian pulse functions -- previously considered by~\citet{Baranov2010}):
	\begin{subequations}
		\label{eq:pulse}
		\begin{align}
		&P_1(t) = \left\{\begin{matrix}
		t_{\rm{las}}^{-1}, \quad t \leq t_{\rm{las}} \\ 
		0, \quad t > t_{\rm{las}} 
		\end{matrix}\right., \\
		&P_2(t) = \frac{2}{t_{\rm{las}}} \cdot \left\{\begin{matrix}
		2t/t_{\rm{las}}, \quad t \leq t_{\rm{las}}/2 \\ 
		2(t-t_{\rm{las}})/t_{\rm{las}}, \quad t_{\rm{las}}/2 < t < t_{\rm{las}} 
		\end{matrix}\right., \\
		&P_3(t)=\frac{5}{\sqrt{\pi}t_{\rm{las}}} \cdot {\rm{exp}} \left( -25 \left[ t/t_{\rm{las}} - 0.5 \right]^2 \right)
		\end{align}
	\end{subequations}
	
	After changing the variables to dimensionless numbers and introducing a computationally efficient sign function~$\mathrm{sgn}(x)$, a quantity~$\Phi(\textrm{Fo}) := P({\rm{Fo}}) \cdot l^2/a$ may be defined, and~Eqs.~\eqref{eq:pulse} can then be re-written:
	\begin{subequations}
		\label{eq:pulse_dimensionless}
		\begin{align}
		&\Phi_{1}({\rm{Fo}}) = 0.5/{\mathrm{Fo}} \times \left[1 + \rm{sgn}\left( {Fo_{\mathrm{las}} - \mathrm{Fo}} \right) \right] \\ 
		&\Phi_{2}(\mathrm{Fo}) = \frac{1}{\mathrm{Fo}_{\mathrm{las}}} \times \left[ 1 + \mathrm{sgn}\left( {\mathrm{Fo}_{\mathrm{las}} - \mathrm{Fo}} \right) \right] \times \nonumber \\
		&\times \left[1 - {\left| \mathrm{Fo_{las}} - 2\mathrm{Fo} \right|}/{\mathrm{Fo}_{\mathrm{las}}} \right] \\ 
		&\Phi_{3}(\mathrm{Fo})=5/{\left ( \sqrt{\pi}\mathrm{Fo}_{\mathrm{las}} \right )} \times \left( -25 \left[ \mathrm{Fo}/\mathrm{Fo}_{\mathrm{las}} - 0.5 \right]^2 \right)
		\end{align}
	\end{subequations}
	
	\subsection{Spatial and temporal domains}
	\label{sect:domain}
	
	The spatial domain of Eqs.~\eqref{eq:boundary_problem_dimensionless} is simply~$y=[0,1]$; the temporal domain depends on the actual data acquisition time~$t_{\rm{max}}$, which is chosen by an algorithm implemented in the instrument software or -- when the latter fails -- by the operator manually. A bad choice of~$t_{\rm{max}}$ can not only result in additional computational costs, but may also lead to a biased estimate of the thermal properties if the accumulated experimental data points are too few, or if the statistics for the temperature rise region of the experimental~$T(\widetilde{t})$ profile are under-represented. In addition, the signal registered at the sample's rear surface often shows a brief excursion during the pulse~$0 \leq t \leq t_{\rm{las}}$ -- an event which cannot be described by a simple model such as that introduced in Sect.~\ref{sect:boundary_problem}. These two aspects can be partially addressed by implementing a data truncation procedure, which would help in establishing a temporal domain suitable for solving the reverse heat conduction problem. The~$\theta({y=1}, \rm{Fo})$ solution of Eqs.~\ref{eq:boundary_problem_dimensionless} and the subsequent conversion of that function to a~$\widehat {T}(t_j)$ dataset~(see~Sect.~\ref{sect:target_function}) need to be defined at a temporal domain~$0 \leq t \leq \widehat{t}_{\rm{max}}$, where~$\widehat{t}_{\rm{max}}$ ensures adequately represented statistics. The~$\widehat{t}_{\rm{max}}$ value may be chosen based on the characteristic thermal transfer time~$\widehat{t}_{\rm{max}} \simeq l^2/a$~(this corresponds to~$\rm{Fo} = 1$ in Eq.~\eqref{eq:fo}), which in turn can be defined in terms of the half-rise time~$t_{1/2}=1.370 l^2 / (\pi^2 a)$~(\citet{heckamn1973}) first introduced by~\citet{Parker1961}. This yields~$\widehat{t}_{\rm{max}} \simeq \pi^2/1.370 t_{1/2} \approx 7.2 t_{1/2}$. Note that the radial heat fluxes have a different characteristic time~$d^2/a$ defined by the sample diameter~$d$. Since typically~$d/l \approx 0.1 - 0.2$, the radiative heat transfer may occur at the same timescale as the radial temperature equilibration. Therefore, estimating the heat losses from a comparison between the experimental and the calculated heating curve at times longer than~$\widehat{t}_{\rm{max}}$ presents an ill-posed problem without explicitly accounting for the radial heat fluxes. Even though the laser beam might not cover the sample completely, these difficulties can be avoided, and the effect of potential signal drift minimized, if the measurement time is kept as short as practically possible -- which is precisely what the above-described truncation procedure does automatically. Thus, the rectangular domain of Eqs.~\eqref{eq:boundary_problem} is defined as: $D = \left( 0 \leq x \leq l, \ 0 \leq t \leq \widehat{t}_{\rm{max}} \right )$, and the domain of~Eqs.~\eqref{eq:boundary_problem_dimensionless} -- as~$\overline{D} = \left( 0 \leq y \leq 1, \ 0 \leq \rm{Fo} \leq 1 \right )$. 
	
	\section{Finite-difference schemes for the heat conduction problem} 
	
	The (semi)analytical solutions considered in~\cite{Cowan1963,CapeLehman1963,ClarkTaylor1975,THERMITUS19974183,LarsonKoyama1967,Azumi1981,LECHNER1993341} with corrections by~\citet{Josell1995} and~\citet{blumm2002improvement} are the ones referenced in the commercial software for the laser flash instruments. Some new corrections have only recently been proposed by~\citet{philipp2019accuracy}. Instead of pursuing the classical approach, it seems more sensible to propose a purely numerical solution. Some steps in this direction have already been taken previously by \citet{Baranov2010}~(with the use of MATLAB) and~\citet{PAVLOV201739}~(FlexPDE). This section describes how to implement finite-difference schemes for the solution of boundary problem described in Sect.~\ref{sect:boundary_problem} without having to use external software packages.
	
	The domain~$\overline{D}$~(Sect.~\ref{sect:domain}) is divided into a uniform grid by introducing the coordinate step size~$h=1/(N-1)$, where~$N$ is the number of individual coordinate points on the grid, and the discrete time step~$\tau = \tau_{\mathrm{F}} h^2$, $\tau_{\mathrm{F}} \in \mathbb{R}$. The grid is used to discretize~$\theta(y,\rm{Fo})$, which becomes~$\theta(\xi_j,\widehat{\rm{Fo}}_m) = \theta_j^{m}$, $j = 0, ..., N-1$, $m = 0,...,m_0$, called the grid function. This section shows different ways to calculate~$\theta_j^{m}$, which can later be converted to~$\widehat{T}(t_j)$~(see Sect.~\ref{sect:conversion}).
	
	\subsection{Application to the boundary problem~(Eq.~\eqref{eq:boundary_problem_dimensionless})}
	\label{sect:scheme_application}
	
	If~$\sigma \in [0,1] \cap \mathbb{R}$, the finite-difference representation of~Eq.~\eqref{eq:he_1d_nodim} on a six-point pattern may be written as~\cite{samarskii2001theory}:
	\begin{eqnarray}
	\label{eq:general_scheme}
	\frac{\theta_j^{m+1} - \theta_j^m}{\tau} = \Lambda \left ( \sigma \theta_j^{m+1} + (1 - \sigma) \theta_j^{m} \right ), \nonumber \\ 
	\ 1 \leq j < N - 1, \ 0 \leq m \leq m_0.
	\end{eqnarray}
	
	Three special cases of~Eq.~\eqref{eq:general_scheme} will be considered: \begin{enumerate*}[label=(\alph*)] \item the fully implicit scheme~($\sigma = 1$), \item the forward-time central-space (FTCS) scheme~($\sigma = 0$), and \item the Crank-Nicolson scheme~($\sigma = 0.5$) \end{enumerate*}. When~$\sigma = 0$, Eqs.~\ref{eq:general_scheme} can be explicitly solved against~$\theta_j^{m+1}$. The solution is straightforward and will be skipped here. If~$\sigma \neq 0$, a sweep algorithm~\citep{samarskii2001theory} may be used to solve the tridiagonal set of linear equations~\eqref{eq:general_scheme}:
	\begin{equation}
	\label{eq:implicit_scheme}
	a_j \theta_{j-1}^{m+1} - b_j \theta_j^{m+1} + c_j \theta_{j+1}^{m+1} = F_j,
	\end{equation}
	where $a_j = c_j = 1/h^2$. For the fully implicit scheme: $b_j = 1/\tau + 2/h^2, \ F_j = - \theta_j^m / \tau$. For the Crank-Nicolson scheme: $b_j = 2/\tau + 2/h^2, \ F_j = - 2\theta_j^m / \tau - \Lambda \theta_j^m$. 
	
	In the sweep algorithm, the following recurrent expression is introduced:
	\begin{equation}
	\label{eq:sweep}
	\theta_j = \overline{\alpha}_{j+1}\theta_{j+1} + \overline{\beta}_{j+1}.
	\end{equation}
	
	Let~$L \phi(\xi_{\alpha}) = \left ( \phi_{\alpha+1} - \phi_{\alpha-1} \right ) / 2h$. The order of approximation for Eq.~\eqref{eq:general_scheme} is~$O(\tau + h^2)$ for either~$\sigma = 0$ or~$\sigma = 1$ and~$O(\tau^2 + h^2)$ for e.g.~$\sigma = 0.5$. On the other hand, the following finite-difference representation of the boundary conditions~(see Eqs.~\eqref{eq:bc_0x_nodim} and \eqref{eq:bc_lx_nodim}) has an order of approximation~$O(h)$:
	\begin{subequations}
		\label{eq:bc_finite}
		\begin{align}
		\label{eq:bc_lower}
		&L \theta_0 = \rm{Bi} \cdot \theta_0 - \Xi, \\
		&L \theta_{N-1} = - \rm{Bi} \cdot \theta_{N-1},
		\end{align} 
	\end{subequations}
	where~$\Xi = \Xi^{m+1}$ is the discretized pulse function defined by substituting~$\mathrm{Fo}_{\mathrm{las}}$, e.g. in Eqs.~\eqref{eq:pulse_dimensionless}, with the discrete pulse width~$\widehat{\mathrm{Fo}}_{\mathrm{las}} = \left \lfloor \frac{ \mathrm{Fo}_{\mathrm{las}} \big / \{ l^2 a^{-1} \} } { \tau } \right \rfloor \cdot \tau$~(square brackets denote the floor function). Note this effectively changes the magnitude of the heat source term in Eq.~\eqref{eq:bc_lower} by~$\sim O(\tau)$, which only slightly affects the maximum heating~(see Eq.~\eqref{eq:delta_tm}); this is compensated automatically when re-scaling the solution to match the experimental heating~$\Delta T_{\mathrm{max}}$ value~(see Sect.~\ref{sect:conversion}). Additionally, the numerical solution described here is valid for any type of the~$\Xi(\widehat{\mathrm{Fo}}_{m+1})$ function, which can be given either in analytic~(e.g. Eqs.~\eqref{eq:pulse_dimensionless}) or tabular form~(i.e., directly measured using a laser power sensor) -- without having to select a fitting function for the pulse shape~(e.g. as done by~\citet{blumm2002improvement}).
	
	To obtain the same order of approximation for the boundary conditions as for~Eq.~\eqref{eq:general_scheme}, a Taylor expansion in the $h$-vicinity of~$\xi = \xi_0$ and $\xi = \xi_{N-1}$ may be used to define virtual nodes~$\xi = \xi_{-1}$ and $\xi = \xi_N$:
	\begin{subequations}
		\label{eq:higher_order_bc}
		\begin{align}
		&\theta_{-1} \simeq \theta_0 - L \theta_0 \cdot h + \Lambda \theta_0 \cdot h^2/2, \\
		&\theta_{N} \simeq \theta_{N-1} + L\theta_{N-1} \cdot h + \Lambda \theta_{N-1} \cdot h^2/2.
		\end{align} 
	\end{subequations}
	
	Noticing that $\Lambda \theta_j^{m+1} = (\theta_j^{m+1} - \theta_j^{m})/\tau \cdot \sigma - \Lambda \theta_j^{m} {(1-\sigma)}/{\sigma}$ and combining~Eqs.~\eqref{eq:bc_finite}--\eqref{eq:higher_order_bc} yields the~$\overline{\alpha}_1$ and $\overline{\beta}_1$ values for Eq.~\eqref{eq:sweep} as well as the grid function value at~$\xi_{N-1}$.
	\begin{enumerate}[label=(\roman*)]
		\item \textit{Fully implicit scheme}~($\sigma = 1$): 
		\begin{subequations}
			\begin{align}
			\label{eq:fi_alpha_lower}
			&\overline{\alpha}_1 = \frac{2 \tau}{h^2 + 2\tau(1 + h\rm{Bi})}, \\ 
			\label{eq:fi_beta_lower}
			&\overline{\beta_1} = \frac{h^2 \theta_0^m + 2h \tau \Xi^{m+1}}{h^2 + 2\tau(1 + h\rm{Bi})}, \\
			\label{eq:fully_implicit_upper_edge}
			&\theta_{N-1}^{m+1} = \frac{h^2 \theta_{N-1}^m + 2 \tau \overline{\beta}_{N-1}}{h^2 + 2 h \tau \rm{Bi} + 2\tau(1 - \overline{\alpha}_{N-1}) }  
			\end{align}
		\end{subequations}
		\item \textit{Crank-Nicolson scheme}~($\sigma = 0.5$):
		\begin{subequations}
			\begin{align}
			\label{eq:cn_alpha_lower}
			&\overline{\alpha}_1 = \frac{\tau}{h^2 + \tau(1 + h\rm{Bi})}, \\
			\label{eq:cn_beta_lower}
			&\overline{\beta_1} = \frac{ (\Xi^{m+1} + \Xi^m) - (\theta_0^m - \theta_1^m )/h + (h/\tau - \rm{Bi}) \theta_0^m }{h/\tau + 1/h + \mathrm{Bi}}, \\
			\label{eq:cn_upper_edge}
			&\theta_{N-1}^{m+1} = - \frac{h \theta_{N-1}^m ( \tau \mathrm{Bi} - h ) - \tau \overline{\beta}_{N-1} + \tau ( \theta_{N-1}^m - \theta_{N-2}^m ) }{h^2 + h \tau \rm{Bi} + \tau \left ( 1 - \overline{\alpha}_{N-1} \right ) }.  
			\end{align}
		\end{subequations}
	\end{enumerate}

	Eqs.~\eqref{eq:fully_implicit_upper_edge} and~\eqref{eq:cn_upper_edge} are used to initiate the calculation of~$\theta_j$, $j = N-2, ..., 0$ with the recurrent expression given by Eq.~\eqref{eq:sweep} where the~$\overline{\alpha}_j$ and~$\overline{\beta}_j$ are calculated using standard expressions~\cite{samarskii2001theory} -- in addition to the~$\overline{\alpha}_1$ and~$\overline{\beta}_1$ values given by Eqs.~\eqref{eq:fi_alpha_lower}, \eqref{eq:fi_beta_lower}, \eqref{eq:cn_alpha_lower}, and \eqref{eq:cn_beta_lower}. When the grid function~$\theta_j^{m+1}$ has been fully calculated for~$j = 0,...,N-1$, the above-described process repeats at the next time step~$m+2$, $m+3$, etc. -- until just above the time limit~$m_{\mathrm{max}} = \rm{Fo}_{\rm{max}} / \tau$~(so that the calculated solution is always defined on a slightly wider temporal domain than the experimental data -- this helps the correct operation of the interpolation procedure as described in Sect.~\ref{sect:target_function}). Further, the number of computed~$\theta^m$ data points is reduced to a pre-set value~$n_\textrm{s}$ with equal spacing between points; hence, only every~$\left \lfloor \mathrm{Fo}_{\rm{max}}/(n_{\mathrm{s}} \tau) \right \rfloor$ point form up the calculated heating curve stored in memory~(square brackets denote the floor function). Finally, the calculated curve is scaled by a factor of~$T_{\rm{max}}/\mathrm{max}{\left ( \theta_{N}^{m} \right )}$~(see Sect.~\ref{sect:conversion}).
	
	\subsection{Cross-verification}
	
\begin{figure}[ht]
	\centering
	\begin{tikzpicture}
	\pgfplotsset{
		legend cell align={left}
	}
	\begin{groupplot}
	[group style={
		group name=cross-verification,
		group size=1 by 2,
		xlabels at=edge bottom,
		ylabels at=edge left,
		yticklabels at=edge left,
	},	
	xmin=0, xmax=1.01,
	ymin=0.0, ymax=1.01,
	width=0.85\columnwidth,
	xlabel={Time, $\mathrm{Fo}$},
	ylabel={Heating, $\Delta\widehat{T}$},
	ylabel near ticks,		
	legend pos=south east,
	legend style={font=\footnotesize},
	]
	\nextgroupplot[title={(\textit{a})}]
	\addplot[color=green,mark=\empty,mark options={scale=0.15}] 
	table[x expr=\thisrowno{4}/0.0706,y expr=\thisrowno{5}/6.964891] {HeatLoss.csv};
	\addlegendentry{$\sigma = 0.5$, $\tau_{\mathrm{F}} \approx 0.25$};
	\addplot[densely dotted, color=blue,mark=\empty,mark options={scale=0.15}] 
	table[x expr=\thisrowno{2}/0.0706,y expr=\thisrowno{3}/6.964891] {HeatLoss.csv};
	\addlegendentry{$\sigma = 1.0$, $\tau_{\mathrm{F}} \approx 0.25$};	
	\addplot[thick, dashed, color=black,mark=\empty,mark options={scale=0.15}] 
	table[x expr=\thisrowno{6},y expr=\thisrowno{7}/0.591865257946954] {HeatLoss.csv};
	\addlegendentry{Ref.~\cite{Josell1995} ($Y = 0.5$)};
	\nextgroupplot[xmax=0.6,title={(\textit{b})}]	
	\addplot[thick,color=red,mark=\empty,mark options={scale=0.15}] 
	table[x expr=\thisrowno{0}/0.0706,y expr=\thisrowno{1}/6.964891] {PulseWidth.csv};
	\addlegendentry{$\mathrm{Fo}_{\mathrm{las}} \approx 0.0004$};
	\addplot[thick, densely dotted, color=blue!25!red,mark=\empty,mark options={scale=0.15}] 
	table[x expr=\thisrowno{2}/0.0706,y expr=\thisrowno{3}/6.964891] {PulseWidth.csv};
	\addlegendentry{$\mathrm{Fo}_{\mathrm{las}} \approx 0.02$};
	\addplot[thick,dashdotted, color=blue!75!red,mark=\empty,mark options={scale=0.15}] 
	table[x expr=\thisrowno{4}/0.0706,y expr=\thisrowno{5}/6.964891] {PulseWidth.csv};
	\addlegendentry{$\mathrm{Fo}_{\mathrm{las}} \approx 0.07$};	
	\end{groupplot}	
	\end{tikzpicture}
	\caption{\label{fig:scheme_verification_heat_loss}The effect of (\textit{a}) heat losses~($l/d = 0.1$) and (\textit{b}) finite pulse width of a rectangular wave~($\sigma = 1.0$, $N=30$). Comparison of the finite-difference solution of the boundary problem in Eq.~\eqref{eq:boundary_problem_dimensionless} using various different schemes~(FTCS, fully implicit and Crank-Nicolson) with the model by~\citet{CapeLehman1963} corrected by~\citet{Josell1995}. The latter model describes the two-dimensional laser heating of a wide, thin disc by an infinitely short laser pulse uniformly covering its front surface; the heating arising from the pulse is followed by cooling through thermal radiation from the front and rear surfaces only characterized by an efficiency factor~$Y = 0.5$~(note that $Y = \mathrm{Bi}$ in current notations). For the numeric solution, the pulse shape has been chosen as rectangular with the pulse duration~$\mathrm{Fo}_{\rm{las}} \approx 4 \times 10^{-4}$.}
	\end{figure}
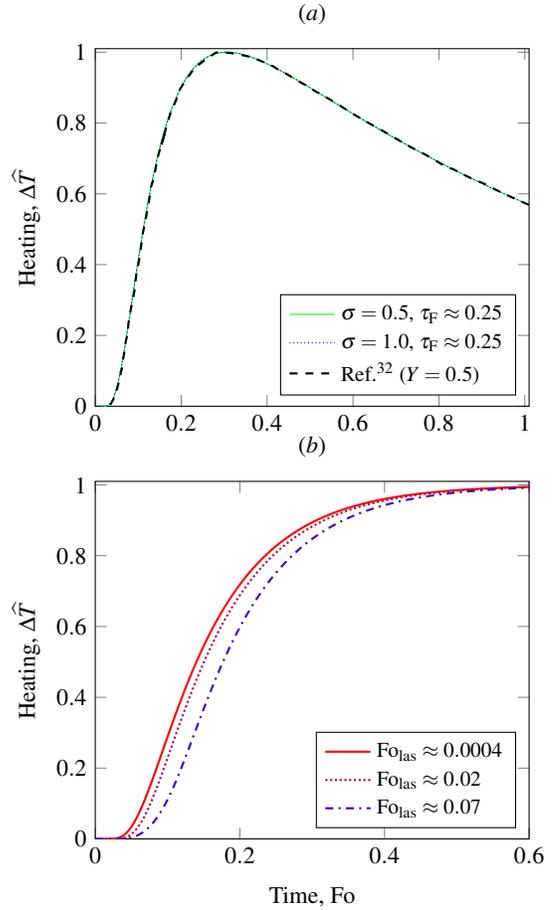

	To verify whether the finite-difference scheme~(Sect.~\ref{sect:scheme_application}) and the boundary problem~(Sect.~\ref{sect:boundary_problem}) have been composed correctly, the numerical solution~$\theta(y=1,\mathrm{Fo})$ calculated using the described procedure is compared to the previously published analytical solutions by~\citet{Parker1961}, \citet{CapeLehman1963} and \citet{Josell1995} for two extreme cases. In both cases the heating of a thin, wide cylinder by an infinitesimal laser pulse is considered. The sample is either~\begin{enumerate*}[label=(\alph*)]
		\item thermally-insulated~($\mathrm{Bi} = 0$) or
		\item cooled down by the radiative heat transfer with an efficiency~$\mathrm{Bi} = 0.5$
	\end{enumerate*}. The results of this comparison are shown in~Figs.~\ref{fig:scheme_verification_adiabatic} and~\ref{fig:scheme_verification_heat_loss}.
	
	\begin{figure}
	\centering
	\begin{tikzpicture}
	\pgfplotsset{
		ylabel near ticks,
		legend cell align={left}
	}
	\begin{axis}[
	xmin=0, xmax=1.0,
	ymin=0.0, ymax=1.05,
	xlabel={Time, $\mathrm{Fo}$},
	ylabel={Heating, $\Delta\widehat{T}$},
	width=0.9\columnwidth,
	set layers=standard,
	legend pos=south east,
	legend style={font=\footnotesize},
	]
	\addplot[color=red,mark=\empty,mark options={scale=0.15}] 
	table[x expr=\thisrowno{0}/0.0706,y expr=\thisrowno{1}/6.964891] {SchemeComparison.csv};
	\addlegendentry{$\sigma = 0.0$, $N = 250$, $\tau_{\mathrm{F}} \approx 0.5$};
	\addplot[color=green,densely dotted,mark=\empty,mark options={scale=0.15}] 
	table[x expr=\thisrowno{6}/0.0706,y expr=\thisrowno{7}/6.964891] {SchemeComparison.csv};
	\addlegendentry{$\sigma = 0.5$, $N = 80$, $\tau_{\mathrm{F}} \approx 0.05$};
	\addplot[color=blue,dashdotted,mark=\empty,mark options={scale=0.15}] 
	table[x expr=\thisrowno{3}/0.0706,y expr=\thisrowno{4}/6.964891] {SchemeComparison.csv};
	\addlegendentry{$\sigma = 1.0$, $N = 80$, $\tau_{\mathrm{F}} \approx 0.05$};
	\addplot[thick,color=black,mark=\empty,dashed] 
	table[x expr=\thisrowno{3}/0.0706,y expr=\thisrowno{5}/6.9648820341184] {SchemeComparison.csv};
	\addlegendentry{\citet{Parker1961}};
	\coordinate (POS1) at (rel axis cs:1.21,0.0);	
	\end{axis}	
	\end{tikzpicture}
	\begin{tikzpicture}
	\begin{groupplot}
	[group style={
		group name=accuracy,
		group size=3 by 2,
		xlabels at=edge bottom,
		vertical sep=40pt,
	},	
	ybar,
	ymin=0,
	width=0.12\textwidth,
	scale only axis,
	x,
	xlabel={Deviation from analytical solution, $\left ( \Delta\widehat{T} - \Delta\widehat{T}_{\mathrm{Parker}} \right) \times 10^{3}$},
	ylabel={Probability density},
	ytick=\empty,
	ylabel near ticks,		
	ylabel style={font=\footnotesize},
	xlabel style={font=\footnotesize},
	ticklabel style = {font=\footnotesize},
	title style={font=\footnotesize},
	]
	\nextgroupplot[xlabel={},ylabel={},title={$\sigma = 0.0$, $N=250$}]
	\addplot +[red!20!black,fill=red!80!white,sharp plot,
	hist={
		density,
		bins=16,
		data min=-5,
		data max=5,
	}   
	] table [sharp plot,y expr = \thisrowno{2}/6.96488193313206E-3 - \thisrowno{1}/6.964891E-3] {SchemeComparison.csv};
	\nextgroupplot[ylabel={},title={$\sigma = 0.5$, $N=80$}]
	\addplot +[green!20!black,fill=green!80!white,sharp plot,
	hist={
		density,
		bins=16,
		data min=-1.0,
		data max=1.0,
	}   
	] table [y expr = \thisrowno{8}/6.9648820341184E-3 - \thisrowno{7}/6.964891E-3] {SchemeComparison.csv};;
	\nextgroupplot[ylabel={},xlabel={},title={$\mathbf{\sigma = 1.0}$, $\mathbf{N=80}$}]	
	\addplot +[
	hist={
		density,
		bins=16,
	}   
	] table [y expr = \thisrowno{5}/6.96487533952232E-3 - \thisrowno{4}/6.964891E-3] {SchemeComparison.csv};
	\nextgroupplot[xlabel={},ylabel={},title={$\sigma = 0.0$,~$N=80$}]
	\addplot +[red!20!black,fill=red!80!white,sharp plot,
	hist={
		density,
		bins=16,
		data min=-5,
		data max=5,
	}   
	] table [sharp plot,y expr = \thisrowno{2}/6.96488221579027E-3 - \thisrowno{1}/6.964891E-3] {SchemeComparison_Small.csv};
	\nextgroupplot[ylabel={},title={$\mathbf{\sigma = 0.5}$, $\mathbf{N=30}$}]
	\addplot +[green!20!black,fill=green!80!white,sharp plot,
	hist={
		density,
		bins=16,
		data min=-1.0,
		data max=1.0,
	}   
	] table [y expr = \thisrowno{8}/6.9648820341184E-3 - \thisrowno{7}/6.964891E-3] {SchemeComparison_Small.csv};;
	\nextgroupplot[ylabel={},xlabel={},title={$\sigma = 1.0$, $N=30$}]	
	\addplot +[
	hist={
		density,
		bins=16,
		data min=-1,
		data max=1,
	}   
	] table [y expr = \thisrowno{5}/6.96487533952232E-3 - \thisrowno{4}/6.964891E-3] {SchemeComparison_Small.csv};	
	\end{groupplot}
	\end{tikzpicture}
	\caption{\label{fig:scheme_verification_adiabatic}Comparison of the finite-difference solution of the boundary problem in Eq.~\eqref{eq:boundary_problem_dimensionless} using various different schemes~(FTCS, fully implicit and Crank-Nicolson) with the analytic solution by~\citet{Parker1961}, which describes the laser heating of a thermally-insulated wide, thin sample by an infinitely short laser pulse~($\mathrm{Bi} = 0.0$). For the numeric solution, the pulse shape has been chosen as rectangular with the pulse duration~$\mathrm{Fo}_{\rm{las}} \approx 1.41 \times 10^{-5}$. The deviation from the analytical solution is shown using probability density histograms of the residuals, plotted for different grid parameters.}
\end{figure}
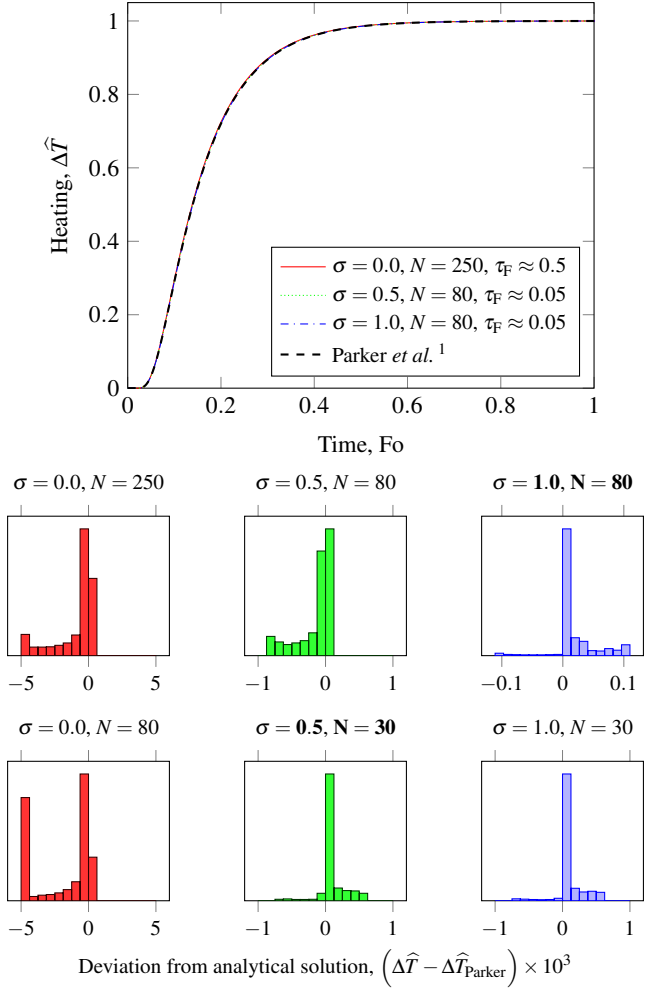

	In each case, cross-verification of the different schemes~($\sigma = 0.0$, $\sigma = 0.5$, $\sigma = 1.0$) was performed. The probability density of residuals~$\Delta\widehat{T} - \Delta\widehat{T}_{\mathrm{Parker}}$ in Fig.~\ref{fig:scheme_verification_adiabatic} has been plotted for different grid parameters~$N$ and~$\tau_{\mathrm{F}}$. The highest accuracy for a very short laser pulse is achieved with the fully implicit scheme~($\sigma = 1.0$) on a dense~($N = 80$, $\tau_{\rm{F}} = 0.05$) grid, where the maximum deviation from the analytical solution~(Fig.~\ref{fig:scheme_verification_adiabatic}) is less than $0.01$\%, and with the Crank-Nicolson scheme~($\sigma = 0.5$) on a loose grid~($N = 30$, $\tau_{\rm{F}} = 6.25 \times 10^{-3}$). Overall, both~$\sigma = 0.0$ and~$\sigma = 1.0$ schemes with an increased order of approximation~$O(\tau + h^2)$ show good numeric stability and high accuracy down to~$N = 15$ and even lower -- compared to a conditionally-stable FTCS scheme with an~$O(h)$ approximation of the boundary conditions. For the second comparison~($\mathrm{Bi} = 0.5$) a very close agreement to the corrected two-dimensional Cape-Lehman model~(\citet{Josell1995}) is shown in~Fig.~\ref{fig:scheme_verification_heat_loss} (a). Fig.~\ref{fig:scheme_verification_heat_loss} (b) shows the solutions for different pulse widths~$\mathrm{Fo}_{\mathrm{las}}$.
	
	\section{Reverse-engineering of the time-temperature profiles to infer thermal properties}
	
	Thermal properties in a laser flash experiment are determined from the minimum of the objective function~$f(\mathbf{S})$ calculated using Eq.~\eqref{eq:ssr} on a curve-to-curve basis. The argument~$\mathbf{S}$ is defined as a variable-size search vector constructed from the values of material-dependant thermal properties~($a/l^2$, $\mathrm{Bi}$ and~$\Delta T_{\rm{max}}$) plus from the parameters~$k_{\mathrm{lin}}$, $T_{\mathrm{lin}}$, and~$t_0$, where~$t_0$ is the time shift between the start of data acquisition and the laser pulse. In cases when noise is large and heat losses are expected to be negligible, to avoid situations where~$\mathrm{Bi}$ may turn negative, it may be explicitly excluded from the search. Thus, the dimension of~$\mathbf{S}$ may vary from two~($a/l^2$ and~$T_{\mathrm{max}}$) to six in the current implementation. 
	
	\subsection{Nonlinear optimization algorithms}
	
	To find~$\mathbf{S}$, a nonlinear optimization algorithm is used to determine the minimum direction~$\mathbf{S}_{\mathrm{min}}$. Once it is found, a linear search determines the optimal magnitude of the step in the~$\mathbf{S}$ direction. These actions are repeated iteratively. If proper algorithms are chosen, with each iteration the objective function~$f(\mathbf{S})$ is brought closer to the global minimum. If the model~(given by Eqs.~\ref{eq:boundary_problem_dimensionless}) is adequate, an unbiased estimate of thermal properties is produced at the global minimum of~$f(\mathbf{S})$. 
	
	\subsubsection{Quasi-Newton method with approximated Hessian}
	\label{sect:quasi-newton}
	
	The calculation of the objective function gradient takes one of the central parts in the direction search routine. If~$\Delta S_i$ is a small variation of the~$i$-th component of the search vector, the associated component of the gradient is~$\Delta f(S_i + \Delta S_i)\Delta S_i$, or more precisely~(\citet{gill1981practical}):
	
	\begin{equation}
	\label{eq:grad}
	g_i = \frac{f(S_i + \Delta S_i) - f(S_i - \Delta S_i)}{2 \Delta S_i},
	\end{equation}
	where a central-difference approximation is used; $f(S_i + \Delta S_i)$ is the value of the objective function calculated at a new search vector~$\overline{\mathbf{S}}$, all components of which are identical to~$\mathbf{S}$ except for the~$i$-th component, which is defined as~$\overline{S}_i = S_i + \Delta S_i$. Eq.~\eqref{eq:grad} introduces a calculation error~$\sim O(\Delta S_i)$.
	
	Consider the following Taylor expansion to the second order of~$f(\mathbf{x}_{k})$ in the~$\gamma_k \mathbf{p}_{k}$ vicinity of~$\mathbf{x}_k = \mathbf{S}_k$:~$f(\mathbf{S}_{k} + \gamma_k \mathbf{p}_{k}) \simeq f(\mathbf{S}_{k}) + \gamma_k \mathbf{g}_{k}^{T}\mathbf{p}_{k} + \mathbf{p}_{k}^{T} \mathbf{G}_k \mathbf{p}_{k}$. The minimum of this quadratic form then corresponds to the Newtonian condition:~$\mathbf{p}_k = -\mathbf{G}_k^{-1} \mathbf{g}_k$. If the Hessian matrix~$\mathbf{G}_k$ is approximated by~$\mathbf{H}_k$, the direction to minimum is then defined as:
	\begin{equation}
	\mathbf{p}_k \approx -\mathbf{H}_k^{-1} \mathbf{g}_k,
	\end{equation}
	where~$\mathbf{H}_k$ is a positive definite matrix containing the curvature information for the objective function. 
	
	A Broyden–Fletcher–Goldfarb–Shanno~(BFGS)~(\citet{gill1981practical}) algorithm is used to calculate~$\mathbf{H}_{k+1}$ at the next iteration~$k+1$ using the gradient value~$\mathbf{g}_k$ and the increment~$\mathbf{u}_k = \mathbf{g}_{k+1} - \mathbf{g}_k$:
	\begin{subequations}
		\label{eq:bfgs}
		\begin{align}
		&\mathbf{H}_{k+1} = \mathbf{H}_{k} + \frac{1}{\mathbf{g}_k^T \mathbf{p}_k} \mathbf{g}_k \mathbf{g}_k^T + \frac{1}{\gamma_k \mathbf{u}_k^T \mathbf{p}_k} \mathbf{u}_k \mathbf{u}_k^T,\\
		&\mathbf{H}_0 = \mathbf{I},
		\end{align}
	\end{subequations}
	where~$\mathbf{I}$ is a~$r \times r$ identity matrix, $r$ is the dimension of the search vector. Thus, the direction to the minimum at the first iteration coincides with the inverted gradient, same as in the gradient descent algorithm, which searches for a stationary point of~$f(\mathbf{S})$. The inverse of the approximated Hessian matrix~$\mathbf{H}_k$ is computed using a recursive procedure based on the Laplace expansion. 
	
	Note Eq.~\eqref{eq:bfgs} is extremely sensitive~\cite{gill1981practical} to the selection of~$\gamma_k$, which is chosen via a linear search algorithm. If the line search fails even by a small margin, $\mathbf{H}_{k+1}^{-1}$ can end up being non-positive definite, which will send the optimizer going uphill instead of downhill.
	
	\subsubsection{Inexact stochastic linear search based on the Wolfe conditions}
	\label{sect:wolfe}
	
	After the vector~$\mathbf{p}_k$ is calculated, the search vector~$\mathbf{S}_{k+1}$ at the next iteration may be expressed as~$\mathbf{S}_{k+1} = \mathbf{S}_{k} + \gamma_k\mathbf{p}_{k}$. If~$\gamma_k \in [0,\gamma_{\mathrm{max}}]$, a linear search may be applied to calculate~$\gamma_k$, which minimizes the objective function~$f(\mathbf{S}_{k} + \gamma_k\mathbf{p}_{k})$. In practice, it is convenient to choose
	\begin{equation}
	\gamma_{\mathrm{max}} = \min \limits_{p_k^i \neq 0} {L^i/p_k^i},
	\end{equation} 
	where~$L^i$ is a safety margin pre-set for~$S^i$ prior to calculation. 
		
	The Wolfe conditions are used to asses whether~$\mathbf{S}_{k+1}$ is likely to be the minimum point. Here the strong Wolfe conditions are considered~(\citet{wolfe1969convergence}, \citet{wolfe1971convergence}):
	\begin{subequations}
		\label{eq:wolfe_all}
		\begin{align}
		\label{eq:wolfe_armijo}
		&f(\mathbf{S}_{k+1}) - (\mathbf{S}_k) \leq c_1 \gamma_k \mathbf{p}_k^T \mathbf{g}_k, \\
		\label{eq:wolfe_strong}
		&| \mathbf{p}_{k}^T \mathbf{g}_{k+1} | \leq c_2 | \mathbf{p}^T_{k} \mathbf{g}_{k} |,
		\end{align}
	\end{subequations}
	where~$c_1 = 0.05$, $c_2 = 0.8$.
	
	The computational procedure utilizing inequalities~\eqref{eq:wolfe_all} starts by generating a random point~$\alpha_k = z$ internal to the~$[a,b]$ segment. The objective function is then calculated at this point, and if~inequality~\eqref{eq:wolfe_armijo} is not satisfied, the segment is then reduced to~$[a,z]$. Otherwise the second condition~(inequality~\eqref{eq:wolfe_strong}) is evaluated. If the latter is satisfied, then~$z$ is considered to be the minimum point, otherwise the computational domain is reduced to~$[z,b]$. The procedure continues while the segment length is greater than~$E_{\rm{lin}}$. 
	
	The algorithm described above works particularly well with the quasi-Newton direction solver.
	
	\subsection{Initial conditions and stopping criteria}
	
	An initial value~$\mathbf{S}_0$ is required and allows to reduce the computational costs if chosen reasonably. Fortunately, this is easy to do once the half-rise time~(Sect.~\ref{sect:conversion}) has been estimated from the experimental data. The initial thermal diffusivity value is determined using the classic solution by~\citet{Parker1961} with the corrected coefficient reported by~\citet{heckamn1973}, \citet{Josell1995} and~\citet{carr2019a}: $a_0 = 1.370 \times l^2/\pi^2 t_{1/2}$. The maximum heating and initial baseline values are determined as described in Sect.~\ref{sect:conversion}, the initial value of the Biot number is set to zero. 
	
	After starting the search, a fixed-size buffer~(the default is eight entries) is filled successively with~$\mathbf{S}_k$ values at each new iteration~$k$; this is complemented by the SSR value~(see Eq.~\eqref{eq:ssr} for a definition). When the buffer is full, the standard deviation of those values~$\delta S{_\omega}$ is calculated for each~$\omega$-th component plus for the SSR. The relative error, calculated as~$\delta S_{\omega} / \left < S_{\omega} \right >$ is then compared to a constant~$E_{\mathrm{gen}}$. If~$\forall \omega \in \mathbb{Z}: \delta S_{\omega} /{\left < S_{\omega} \right >} \leq E_\mathrm{gen}$, the search completes normally. Otherwise, the buffer is cleared and the search continues until the criterion is finally satisfied or if the iteration limit is reached.   
	
	\subsection{Example applications and convergence tests}
	
	A few previously measured time-temperature profiles were selected to represent a gradual deviation from the perfect experimental conditions. These deviations are introduced by different experimental factors, which affect either the perceived temperature rise or the actual heat transfer in the sample. The pre-selected data is then reverse-engineered following the procedure described above to produce an optimal~$\mathbf{S}_k$. The performance of this procedure under real-world conditions is judged and the convergence is tested.
	
	The raw data was collected using two experimental installations:
	\begin{enumerate}[label=(\alph*)] \item the Kvant laser flash analyzer designed at the Moscow Engineering Physics Institute equipped with two temperature detection capabilities: \begin{enumerate*}[label=(\textit{\roman*})] \item a thermocouple~(pre-welded onto the rear surface of the sample coated with platinum black prior to each experiment) connected to a scalable amplifier with automatic constant-voltage subtraction providing rapid detection of heating with an error of less than~$2.5 \times 10^{-3}$~K and \item an in-house designed InGaAs pyrometer~(field of view~$d_{\mathrm{FOV}} = 6$~mm)\end{enumerate*}. A combination of a rotary pump and a diffusion pump ensure high vacuum~($0.01$~Pa or~$10^{-4}$~mbar). An additional capability of controlling the oxygen partial pressure is provided by a solid-electrolyte galvanic cell; the oxygen partial pressure can be changed on-the-fly. A ruby laser~(wavelength~$694.3$~nm) delivers~$1.5$~ms fixed-width pulses with an energy of~$5-7$~J per pulse; 
		
	\item the Linseis Culham LFA, a prototype based on the Linseis LFA 1600 instrument, modified by the manufacturer to allow integration in a research room in the Materials Research Facility~(MRF) at UKAEA for testing of mildly-radioactive samples. The furnace ambient temperature is measured using a low-resolution pyrometer, while a PbSe-based Peltier-cooled detector is used to register the rear-surface heating of the sample. A simple rotary pump is capable of pumping the chamber down to~$0.1 - 0.01$~mbar. The pulse width and pulse energy~(up to~$31$~J) of the Nd:YAG laser~(wavelength~$1064$~nm), as well as the the detector gain and aperture can be changed by the operator during the experimental run. \end{enumerate} 
	The materials under study were the following: a sintered sample of nearly-stoichiometric uranium dioxide~($l=1.7118$~mm, $d=10.061$~mm) with a porosity of about~$8.5$~\%, an MPG-6 graphite sample~($l=2.9302$~mm, $d=9.9654$~mm), and two 10-mm Zr-1\%Nb E110 alloy discs cut by electrical discharge machining~($l = 0.414$~mm and $l = 0.199$~mm). The raw heating curve analyzed here for the uranium dioxide pellet was reported previously~(e.g., \citet{Lunev2011high}). The measurements of the E110 alloy samples were conducted with the Linseis Culham LFA, graphite-spraying both surfaces of the samples prior to tests with a Graphit 33 Contact Chemie™ coating. 
	
	Thermal properties are reverse-engineered from experimental data using the heat transfer model given by Eqs.~\eqref{eq:boundary_problem_dimensionless}. A quasi-Newton direction search (Sect.~\ref{sect:quasi-newton}) with a linear stochastic search algorithm based on the Wolfe conditions (Sect.~\ref{sect:wolfe}) was applied to reach the minimum of the objective function~(gradient resolution~$\Delta S_i/S_i = 10^{-4}$, linear search error~$E_{\mathrm{lin}} = 10^{-7}$, overall search error~$E_{\mathrm{gen}}=10^{-3}$). The model solution was calculated using a fully implicit~($\sigma = 1.0$) difference scheme~(see Eq.~\eqref{eq:general_scheme}) with the default grid settings~$N = 30, \ \tau_{\mathrm{F}} = 0.25$~(total number of points for the model curve~$n_{\mathrm{s}} = 100$).
	
	\subsubsection{Uniform low noise}	
	
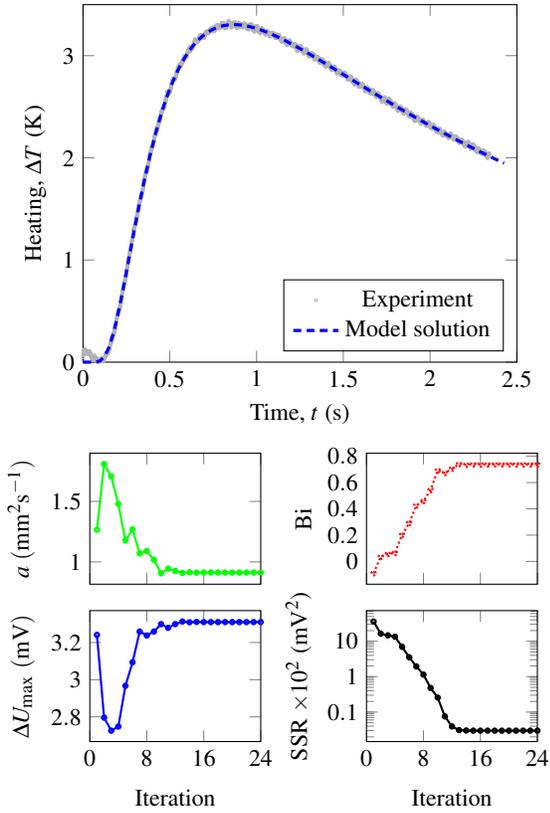
\begin{figure}[!ht]
		\centering
		\begin{tabular}{cc}
			\begin{tikzpicture}
			\pgfplotsset{
				xmin=0, xmax=2.5,
				ymin=0, ymax = 3.5,
				xlabel={Time, $t$ (s)},
				ylabel={Heating, $\Delta T$~(K)},
				width=0.85\columnwidth,
				ylabel near ticks,
				xlabel near ticks,
				legend style={font=\small},
			}
			\begin{axis}[set layers=standard,legend pos=south east]
			\addplot[color=gray!60!white,mark=o,only marks,mark options={scale=0.25}] 
			table[x expr=\thisrowno{0},y expr=\thisrowno{1}] {UO2_Kvant_1829K.csv};
			\addlegendentry{Experiment};
			\addplot[densely dashed, very thick,color=blue,mark=\empty,on layer={axis foreground}] 
			table[x expr=\thisrowno{2},y expr=\thisrowno{3}] {UO2_Kvant_1829K.csv};
			\addlegendentry{Model solution};
			\end{axis}
			\end{tikzpicture}
			\\
			\begin{tikzpicture}
			\begin{groupplot}
			[group style={
				group name=convergence_low_noise,
				group size=2 by 2,
				xlabels at=edge bottom,
				xticklabels at=edge bottom,
				horizontal sep=40pt,
				vertical sep=10pt,			
			},	
			xmin=0, xmax =24,
			width=0.215\textwidth,
			xlabel={Iteration},
			ylabel near ticks,
			xtick distance = 8,
			ylabel style={font=\small},
			xlabel style={font=\small},
			ticklabel style = {font=\small},
			]
			\nextgroupplot[ylabel={$a \ (\mathrm{mm}^2 \mathrm{s}^{-1})$}] 
			\addplot[thick, color=green,mark=o,mark options={scale=0.4}] 
			table[x expr=\thisrowno{0},y expr=\thisrowno{7}*1E6] {UO2_Wolfe};	
			\nextgroupplot[ylabel={Bi},yticklabel pos=left,ytick pos=right]
			\addplot[thick, densely dotted,color=red,mark=o,mark options={scale=0.4}] 
			table[x expr=\thisrowno{0},y expr=\thisrowno{4}] {UO2_Wolfe};	
			\nextgroupplot[ylabel={$\Delta U_{\mathrm{max}} \ (\mathrm{mV})$}]
			\addplot[thick, color=blue,mark=o,mark options={scale=0.4}] 
			table[x expr=\thisrowno{0},y expr=\thisrowno{5}] {UO2_Wolfe};
			\nextgroupplot[ylabel=SSR~$\times 10^{2}$~($\mathrm{mV}^2$),ymode=log,log ticks with fixed point]
			\addplot[thick, color=black,mark=o,mark options={scale=0.4}] 
			table[x expr=\thisrowno{0},y expr=\thisrowno{6}/100] {UO2_Wolfe};	
			\end{groupplot}	
			\end{tikzpicture}
		\end{tabular}
		\caption{\label{fig:example_uo2}An example run of the reverse-engineering procedure for an experimental time-temperature profile measured with an InGaAs detector~(total number of data points~$n_{\mathrm{exp}} = 4,800$) of a~$91.5$~\% dense UO$_2$ sample at~$T_0 = 1829 \ \mathrm{K}$, showing the final heating curve corresponding to the optimized set of model parameters and the components of the search vector, fully converged after~24 iterations. Note the small plateau on the SSR plot, which is likely attributed to a local minimum of the objective function.}
	\end{figure}

	Fig.~\ref{fig:example_uo2} shows a heating curve for a nearly-stoichiometric UO$_2$ sample at a test temperature~$T_0 = 1829.7$~K. Due to a poor thermal conductivity, which is especially low at high temperatures, the characteristic heat conduction time is high. Conversely, the heat losses due to radiation are huge. The heat transfer model~(defined by Eqs.~\eqref{eq:boundary_problem_dimensionless}) perfectly describes the experimental data, except for the initial segment, which is perhaps attributed to the special optical properties of the material~(\citet{Lunev2011high}). Since the baseline was accurately determined by the instrument software, only the thermal diffusivity~$a$, the Biot number~$\mathrm{Bi}$, and the maximum temperature~$\Delta T_{\mathrm{max}}$ were included in the search vector. Excellent convergence is obtained already after~$16$ iterations, with the mean deviation per point~$|{\Delta \widehat{T}_i - \Delta T_i}| \approx 1.5~\times 10^{-2}$~K. Interestingly, the first iteration~(same as for the gradient descent method) results in a shift of the search vector to a different local minimum, but the numerical procedure quickly escapes it and gets to the right track to the global minimum. 
	
	\subsubsection{Synchronization errors}	
	
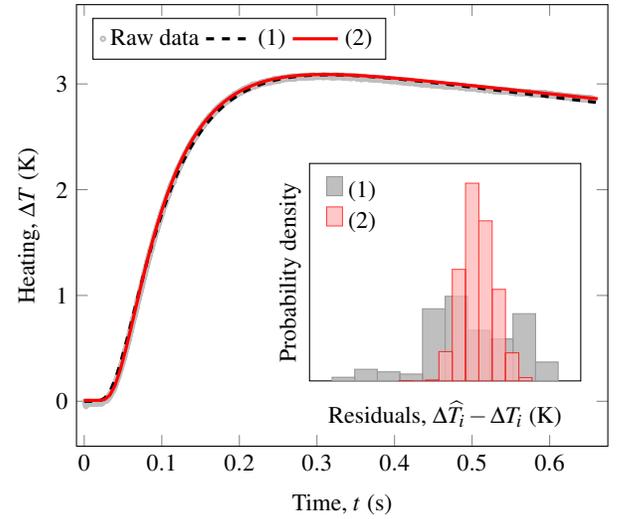
\begin{figure}
	\centering
	\begin{tikzpicture}
	\begin{axis}[ymax = 3.75,legend columns=-1,legend pos = north west,width=1.0\columnwidth,ylabel near ticks,
	set layers=standard,xlabel={Time, $t$ (s)},
	ylabel={Heating, $\Delta T$~(K)},xmin=-0.01,xmax=0.675]
	\addplot[thick,color=black!25!white,mark=o,only marks,mark options={scale=0.4}]  
	table[x expr=\thisrowno{0},y expr=\thisrowno{1}] {RawData_0.csv};
	\addlegendentry{Raw data};
	\addplot[very thick,dashed,color=black,mark=\empty,on layer={axis foreground}] 
	table[x expr=\thisrowno{0},y expr=\thisrowno{1}] {Solution_01.csv};
	\addlegendentry{(1)};
	\addplot[very thick,color=red,mark=\empty,on layer={axis foreground}] 
	table[x expr=\thisrowno{0}+0.006167390292808788,y expr=\thisrowno{1} + 0.00843225986167542] {Solution_21.csv};
	\addlegendentry{(2)};
	\coordinate (inset1) at (rel axis cs:0.95,0.025);
	\end{axis}
	\begin{axis}
	[at={(inset1)},xlabel near ticks,ylabel near ticks,anchor={outer south east},ticklabel style = {font=\footnotesize},ymin=0,ybar,legend pos=north west,legend style={draw=none},ylabel={\small{Probability density}},xtick=\empty,ytick=\empty,xlabel={\small{Residuals, ${\Delta \widehat{T}_i - \Delta T_i}$ (K)}},width=0.6\columnwidth]
	\addplot +[
	hist={
		density,
	},fill=black!25!white,draw=black!40!white
	] table [y index=3] {Solution_0.csv};
	\addlegendentry{(1)};
	\addplot +[
	hist={
		density,		
	},   
	opacity=0.725  
	] table [y index=3] {Solution_2.csv};
	\addlegendentry{(2)};
	\end{axis} 
	\end{tikzpicture}
	\caption{\label{fig:pulse_monitor_fault}An illustration of the systematic error caused by a loss of synchronization in timing of the data acquisition and the laser pulse. The reverse heat problem is first solved using three variables~($a$,~$T_{\mathrm{max}}$, and~$\mathrm{Bi}$), giving rise to the heating curve~$(1)$, which corresponds to~$R^2 = 0.99943$ and a thermal diffusivity value~$a = a_0$. The distribution of residuals in this case differs from the ideal Gaussian function with a zero mean. When the time shift~$t_0$ and the baseline intercept~$T_{\mathrm{lin}}$ are included in the search, the heating curve~$(2)$ is obtained as a result, which yields a Gaussian-type distribution of residuals and a markedly higher~$R^2 = 0.99993$ value~($(a-a_0)/a_0 = 9.73 \% $).}  
	\end{figure}

	When the data acquisition from the detector is slightly ahead or behind the laser pulse, the standard processing procedure will show systematic errors. This is illustrated in Fig.~\ref{fig:pulse_monitor_fault} for an MPG-6 graphite sample~(measurements with an InGaAs detector / Kvant). The loss of synchronization error can be easily identified by looking at the distribution of residuals~(inset in Fig.~\ref{fig:pulse_monitor_fault}). As per the central limit theorem, the latter should follow a Gaussian distribution if only random noise is present in raw data. A deviation from this rule means a source of systematic error needs to be accounted for; in this case, this is done by including the time shift~$t_0$ and the baseline intercept~$T_{\mathrm{lin}}$ in the search vector. The solution to the reverse problem results in a near-perfect distribution of the residuals, meaning the systematic error has now been completely excluded and the results are therefore more accurate.
	
	\subsubsection{Non-uniform high noise}
	
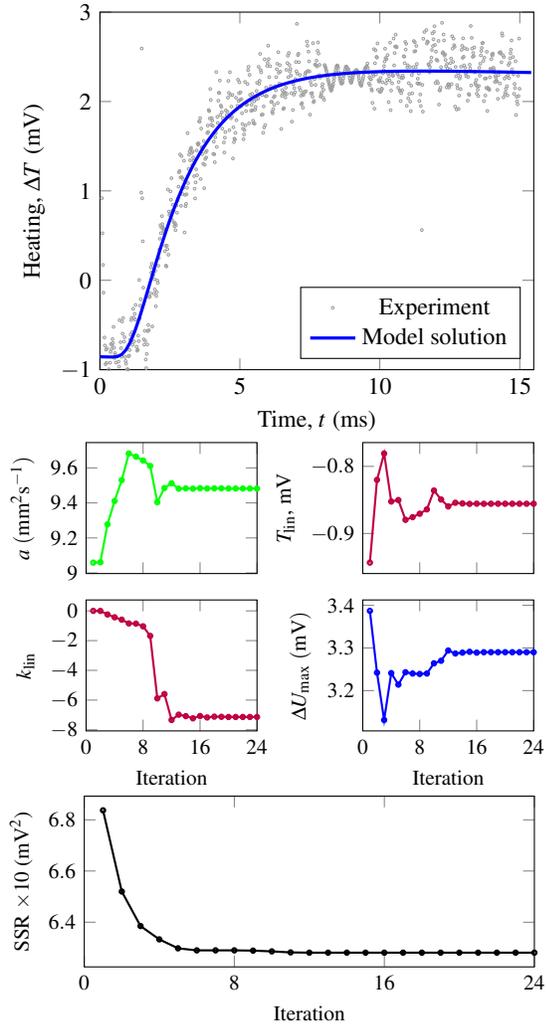
\begin{figure}[!ht]
		\centering
		\begin{tikzpicture}
		\pgfplotsset{
			xmin=0, xmax=15.5,
			ymin=-1.0,ymax=3.0,
			ytick distance = 1.0,
			xlabel={Time, $t$ (ms)},
			ylabel={Heating, $\Delta T$~(mV)},
			width=0.85\columnwidth,
			ylabel near ticks,
			xlabel near ticks,
			legend style={font=\small},		
		}
		\begin{axis}[set layers=standard,legend pos=south east]
		\addplot[color=gray!75!white,mark=o,only marks,mark options={scale=0.25}] 
		table[x expr=\thisrowno{0}*1000,y expr=\thisrowno{1}] {Zr1Nb_1024K_Linseis.csv};
		\addlegendentry{Experiment};
		\addplot[very thick,color=blue,mark=\empty,on layer={axis foreground}] 
		table[x expr=\thisrowno{2}*1000,y expr=\thisrowno{3} + (-8.557e-01 -7.136e+00*\thisrowno{2})] {Zr1Nb_1024K_Linseis.csv};
		\addlegendentry{Model solution};
		\end{axis}		
		\end{tikzpicture}
		\\	
		\begin{tikzpicture}
		\begin{groupplot}
		[group style={
			group name=convergence_high_noise,
			group size=2 by 2,
			xlabels at=edge bottom,
			xticklabels at=edge bottom,
			horizontal sep=40pt,
			vertical sep=10pt,			
		},	
		xmin=0, xmax =24,
		width=0.215\textwidth,
		xlabel={Iteration},
		ylabel near ticks,
		xlabel near ticks,
		xtick distance = 8,
		ylabel style={font=\footnotesize},
		xlabel style={font=\footnotesize},
		ticklabel style = {font=\footnotesize},
		title style={font=\footnotesize},			
		]
		\nextgroupplot[ylabel={$a \ (\mathrm{mm}^2 \mathrm{s}^{-1})$}] 
		\addplot[thick, color=green,mark=o,mark options={scale=0.4}] 
		table[x expr=\thisrowno{0},y expr=\thisrowno{7}*1E6] {ZrNb_1024K_Linseis};	
		\nextgroupplot[ylabel={$T_{\rm{lin}}$, mV},ytick distance=0.1]
		\addplot[thick, color=purple,mark=o,mark options={scale=0.4}] 
		table[x expr=\thisrowno{0},y expr=\thisrowno{1}] {ZrNb_1024K_Linseis};	
		\nextgroupplot[ylabel={$k_{\rm{lin}}$}]
		\addplot[thick, color=purple,mark=o,mark options={scale=0.4}] 
		table[x expr=\thisrowno{0},y expr=\thisrowno{2}] {ZrNb_1024K_Linseis};	
		\nextgroupplot[ylabel={$\Delta U_{\mathrm{max}} \ (\mathrm{mV})$}]
		\addplot[thick, color=blue,mark=o,mark options={scale=0.4}] 
		table[x expr=\thisrowno{0},y expr=\thisrowno{5}] {ZrNb_1024K_Linseis};
		\end{groupplot}	
		\end{tikzpicture}
		\\
		\begin{tikzpicture}
		\begin{axis}[
		xmin=0, xmax =24,
		width=0.875\columnwidth,
		height=0.215\textwidth,
		xlabel={Iteration},
		ylabel near ticks,
		xlabel near ticks,
		xtick distance = 8,
		ylabel style={font=\footnotesize},
		xlabel style={font=\footnotesize},
		ticklabel style = {font=\footnotesize},
		title style={font=\footnotesize},
		ylabel=SSR~$\times 10$~($\mathrm{mV}^2$),
		]
		\addplot[thick, color=black,mark=o,mark options={scale=0.4}] 
		table[x expr=\thisrowno{0},y expr=\thisrowno{6}/10] {ZrNb_1024K_Linseis};	
		\end{axis}
		\end{tikzpicture}
		\caption{\label{fig:example_zr1nb}An example run of the reverse-engineering procedure for an experimental time-temperature profile with a high non-uniform noise~(likely consisting of several noise harmonics) measured with a PbSe detector~(total number of data points~$n_{\mathrm{exp}} = 864$) of a graphite-coated~$l=0.414$~mm E110 alloy sample at~$T_0 = 1024 \ \mathrm{K}$, showing the final heating curve corresponding to the optimized set of model parameters and the components of the search vector, fully converged after~24 iterations.}
	\end{figure}

	Fig.~\ref{fig:example_zr1nb} shows a heating curve for a~$l=0.414$~mm E110 alloy sample at a test temperature~$T_0 = 1024$~K. The material tested with the Linseis Culham LFA is a fair thermal conductor and the sample is sufficiently thin so that the heat wave reaches the rear surface of the sample relatively fast. Because of this, the heat losses are negligible at this and even higher temperature, hence they can be excluded from the search vector. On the other hand, the baseline~(initially defined at~$k_{\mathrm{lin}} = 0$ as explained in Sect.~\ref{sect:conversion} needs to be adjusted due to the inaccurate baseline estimation based on the under-represented statistics at~$t < 0$ from noisy data. Same as previously, full convergence is obtained after~$24$ iterations, with the average data spread of~$\approx 0.25$~mV. The search for the optimal baseline slope seems to be taking the longest time.
	
	\subsubsection{Pathological data}
	
\begin{figure}
		\centering
		\begin{tikzpicture}
		\pgfplotsset{
			xmin=0,xmax=4.75,
			ymin=-8.1, ymax=4,
			xlabel={Time, $t$ (ms)},
			ylabel={Heating, $\Delta T$~(mV)},
			width=0.85\columnwidth,
			ylabel near ticks,
			ytick distance = 4.0,
			xlabel near ticks,
			legend style={font=\small},				
		}
		\begin{axis}[set layers=standard,legend pos=south east]
		\addplot[color=gray!75!white,mark=o,only marks,mark options={scale=0.25}] 
		table[x expr=\thisrowno{0}*1000,y expr=\thisrowno{1}] {Zr1Nb_673K_Linseis_Pathological.csv};
		\addlegendentry{Experiment};
		\addplot[very thick,color=blue,mark=\empty,on layer={axis foreground}] 
		table[x expr=\thisrowno{2}*1000,y expr=\thisrowno{3} + (-8.082e+00 + 2.567E02*\thisrowno{2})] {Zr1Nb_673K_Linseis_Pathological.csv};
		\addlegendentry{Model solution};
		\end{axis}	
		\end{tikzpicture}
		\\
		\begin{tikzpicture}		
		\begin{groupplot}
		[group style={
			group name=convergence_pathological,
			group size=2 by 2,
			xlabels at=edge bottom,
			xticklabels at=edge bottom,
			horizontal sep=40pt,
			vertical sep=10pt,			
		},	
		xmin=0, xmax =24,
		width=0.215\textwidth,
		xlabel={Iteration},
		xlabel near ticks,
		ylabel near ticks,
		xtick distance = 8,
		ylabel style={font=\footnotesize},
		xlabel style={font=\footnotesize},
		ticklabel style = {font=\footnotesize},
		]
		\nextgroupplot[ylabel={$a \ (\mathrm{mm}^2 \mathrm{s}^{-1})$}] 
		\addplot[thick, color=green,mark=o,mark options={scale=0.4}] 
		table[x expr=\thisrowno{0},y expr=\thisrowno{7}*1E6] {Zr1Nb_Pathological_1};	
		\nextgroupplot[ylabel={$k_{\rm{lin}} \times 10^{2}$},yticklabel pos=left]
		\addplot[thick, color=purple,mark=o,mark options={scale=0.4}] 
		table[x expr=\thisrowno{0},y expr=\thisrowno{2}/100] {Zr1Nb_Pathological_1};	
		\nextgroupplot[ylabel={$\Delta U_{\mathrm{max}} \ (\mathrm{mV})$}]
		\addplot[thick, color=blue,mark=o,mark options={scale=0.4}] 
		table[x expr=\thisrowno{0},y expr=\thisrowno{5}] {Zr1Nb_Pathological_1};
		\nextgroupplot[ylabel=SSR~$\times 10^2$~($\mathrm{mV}^2$),yticklabel pos=left]
		\addplot[thick, color=black,mark=o,mark options={scale=0.4}] 
		table[x expr=\thisrowno{0},y expr=\thisrowno{6}/100] {Zr1Nb_Pathological_1};	
		\end{groupplot}	
		\end{tikzpicture}
		\caption{\label{fig:example_pathological_1}An example run of the reverse-engineering procedure for an experimental time-temperature profile with a highly-inaccurate temperature measurement combined with a detector signal drift measured with a PbSe detector~(total number of data points~$n_{\mathrm{exp}} = 1,050$) of a graphite-coated~$l=0.199$~mm E110 alloy sample at~$T_0 = 674 \ \mathrm{K}$, showing the final heating curve corresponding to the optimized set of model parameters and the components of the search vector, fully converged after~24 iterations.}
	\end{figure}
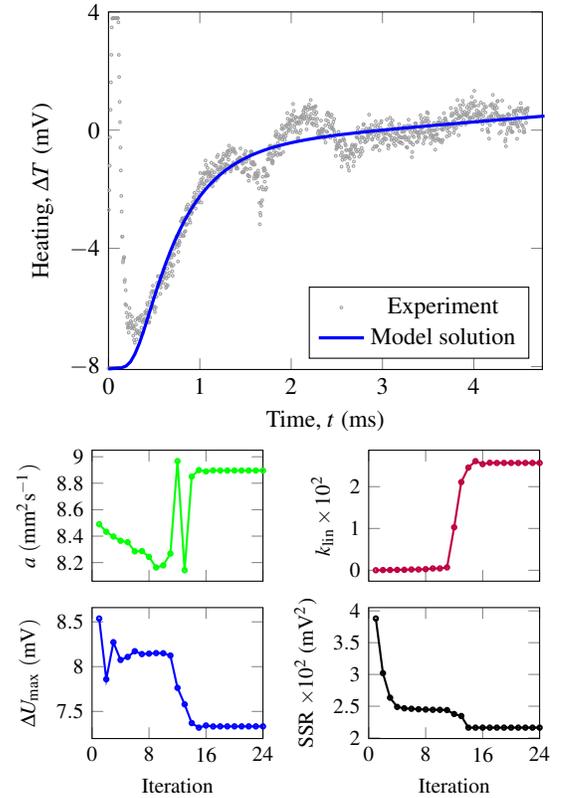

	Fig.~\ref{fig:example_pathological_1} shows a heating curve for a~$l=0.199$~mm E110 alloy sample at a test temperature~$T_0 = 1024$~K measured with the Linseis Culham LFA. The low thickness of the sample leads to rapid heat conduction, with the acquisition time less than~$5$~ms. At this time the baseline intercept estimated at~$t < 0$ seemed to be sufficiently accurate as it was measured over a longer time interval than the actual heating curve. On the other hand, the data obviously showed either a detector signal drift or an ambient temperature instability, which needed to be corrected for by adjusting the baseline slope. Additionally, the data in Fig.~\ref{fig:example_pathological_1} shows anomalies with pronounced low-frequency, high-amplitude noise, which distorts the heating curve so that it becomes nearly impossible to asses its true shape. However, applying the nonlinear optimization procedure described in this section allows reaching a global minimum of the objective function in less than~$24$ iterations, with an average data spread of~$\approx 0.45$~mV. 
	
	\section{Cross-validation with external reference and commercial software}
	
\begin{figure}
		\centering	
		\begin{tikzpicture}
		\pgfplotsset{
			xmin=400,xmax=2400,
			ymin=23, ymax=63,
			xlabel={Temperature, $T_0$ (K)},
			ylabel={Thermal diffusivity, $a \ (\mathrm{mm}^2\mathrm{s}^{-1})$},
			width=1.0\columnwidth,
			ylabel near ticks,
			ytick distance = 10,
			xtick distance = 400,
			grid=both,
			major grid style={densely dotted,line width=.2pt,draw=gray!75},
		}
		\begin{axis}[legend pos=north east,mark options={scale=0.5}]
		\addplot[smooth,thick,color=black,mark=\empty] 
		table[x expr=\thisrowno{0},y expr=\thisrowno{1}] {Netzsch_T_a.txt};
		\addlegendentry{Netzsch reference};
		\addplot+[ultra thick,color=blue,mark=square*,only marks,
		error bars/.cd,y dir=both,x dir = both,x explicit,y explicit,error bar style={line width=1.0pt,solid}] 
		table[x expr=\thisrowno{2},y expr=\thisrowno{4}, y error expr=\thisrowno{5}] {AverageResults_3.csv};
		\addlegendentry{PULsE};	
		\addplot[thick,color=red,mark=x,only marks,mark options={scale=0.75}]  
		table[x expr=\thisrowno{1}+273,y expr=\thisrowno{2}*100]
		{Linseis_model};
		\addlegendentry{Linseis Aprosoft};	
		\end{axis}		
		\end{tikzpicture}
		\caption{\label{fig:validation_summary}Cross-validation results for PULsE~(manual processing of each individual curve) and Linseis AproSoft~(batch processing) using the same dataset acquired with a PbSe detector / Linseis Culham LFA on a standard tungsten sample supplied by Netzsch, pre-coated with graphite. The solid line is plotted using the reference values provided in the documentation for the standard sample~(no information on the error values or the type of detector is available).}
	\end{figure}
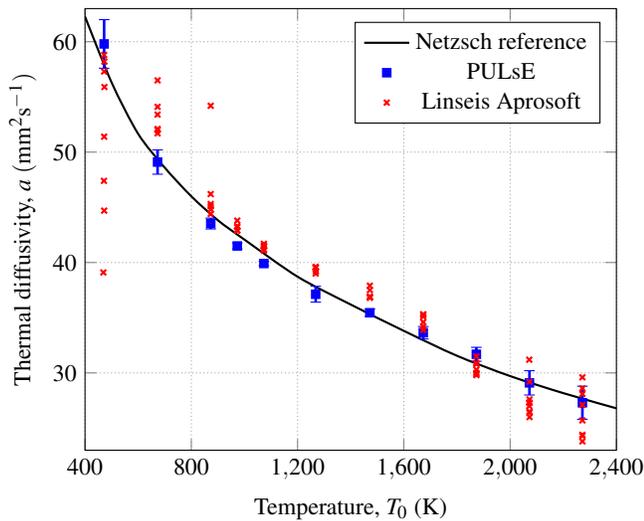

	An attested reference tungsten sample~($l=2.034$~mm, $d=9.88$~mm) was purchased from Netzsch to conduct an independent validation study. The sample arrived with a printed copy of a reference table listing the pre-measured thermal diffusivity values in accordance with the ASTM standard~\cite{astm2013standard}. The sample was then prepared for laser flash measurements following the manufacturer and ASTM recommendations by coating both sides with a thin layer of high-emissivity carbon. As previously, a Graphit 33 Contact Chemie™ spray was used on the sample pre-heated at~$100$~\celsius. The coated tungsten sample had its thickness measured with a micrometer and then was placed in a graphite sample holder on top of a graphite thermal shield of the Linseis Culham LFA. The system was evacuated to a pressure below~$10^{-1}$~mbar and filled with a Grade Zero argon gas. This process was repeated three times, after which the gas flow was set to~$8 \ \mathrm{l}/\mathrm{h}$. The heating curves were measured in a range of temperatures~$T_0 = 473 - 2273$~K~(heating and cooling rates were $20\celsius /\mathrm{min}$) with a PbSe detector. At each temperature, the detector parameters~(gain, iris, and acquisition time), as well as the laser power and pulse duration, were changed manually by the operator to deliver the best signal-to-noise ratio and recorded in a metadata file separately. Measurements and data processing using the ``Combined model'' and baseline subtraction were controlled from the commercial Linseis Aprosoft v1.06 software adapted specifically for the Linseis Culham LFA. Default settings were used for the temperature and detector current stability controls, and a constant delay of~$90$~s was used between shots. The resulting thermal diffusivity data from the first run~(to reduce the formation of tungsten carbide) is plotted in~Fig.~\ref{fig:validation_summary}.
	
	The recorded detector signal exported in ASCII-format and metadata files prepared by the operator were used as input when running PULsE. Again, as previously, a quasi-Newton direction search (Sect.~\ref{sect:quasi-newton}) with a linear stochastic search algorithm based on the Wolfe conditions~(Sect.~\ref{sect:wolfe}) were adopted for data treatment~(a gradient resolution~$\Delta S_i/S_i = 10^{-4}$, a linear search error~$E_{\mathrm{lin}} = 10^{-7}$ and a global search error~$E_{\mathrm{gen}}=10^{-3}$). A fully implicit~($\sigma = 1.0$) difference scheme was used to calculate~$\widehat{T}(t)$~(see Eq.~\eqref{eq:general_scheme}) with the default grid settings~$N = 30, \ \tau_{\mathrm{F}} = 0.25$~(total number of points for the model curve~$n_{\mathrm{s}} = 100$). 
	
\begin{figure*}[!ht]
		\centering
		\begin{tikzpicture}
		\begin{groupplot}
		[group style={
			group name=convergence,
			group size=2 by 1,
			xlabels at=edge bottom,
			horizontal sep=50pt,
		},
		width=0.65\columnwidth,
		ylabel near ticks,
		set layers=standard,
		xlabel={Time, $t$ (ms)},
		ylabel={Heating, $\Delta T$~(mV)},
		]
		\nextgroupplot[xmin=0,xmax=90,ymin=-16,ymax=2.5,legend pos=south east,title={(\textit{a}) $T_0=469$~K}]
		\addplot[very thick,color=red,mark=\empty,on layer={axis foreground}] 
		table[x expr=\thisrowno{2},y expr=\thisrowno{3}] {Linseis_Solution_201};
		\addplot[domain=0:90,restrict x to domain=0:90,very thick,densely dotted,color=blue,mark=\empty,on layer={axis foreground}] 
		table[x expr=\thisrowno{0}*1000,y expr=\thisrowno{1}+(-15.999 -16.7732*\thisrowno{0})] {PULSE_Solution_201};
		\addplot[color=black!25!white,mark=o,only marks,mark options={scale=0.5}]  
		table[x expr=\thisrowno{0}*1000,y expr=\thisrowno{1}]
		{Experimental_Data_201}
		node[pos=0.085,pin={[gray,very thick]right:Detection limit}]{};
		\nextgroupplot[legend pos=south east,xmin=0,xmax=61,ymin=-0.3,ymax=3.25,,title={(\textit{b}) $T_0=973$~K}]
		\addplot[thick,color=red,mark=\empty,on layer={axis foreground}] 
		table[x expr=\thisrowno{2},y expr=\thisrowno{3}] {Linseis_Solution_224};
		\addlegendentry{Linseis};
		\addplot[very thick,densely dotted,color=blue,mark=\empty,on layer={axis foreground}] 
		table[x expr=\thisrowno{0}*1000,y expr=\thisrowno{1}+(-0.21321)] {PULSE_Solution_224};
		\addlegendentry{PULsE};
		\addplot[thick,color=black!25!white,mark=o,only marks,mark options={scale=0.75}]  
		table[x expr=\thisrowno{0}*1000,y expr=\thisrowno{1}]
		{Experimental_Data_224};	
		\end{groupplot}		
		\end{tikzpicture}
		\caption[foo_bar]{\label{fig:validation_example}Example benchmarking of PULsE on two datasets: \begin{enumerate*}[label=(\alph*)] \item an experiment at~$T_0 = 469~K$. Due to a detector failure, the signal saturated at~$-10$~mV for the first~$\approx 12$~ms after the laser shot. PULsE was capable of correcting for that error after manually limiting the time range to exclude problematic values, resulting in a value of thermal diffusivity~$a=56.03$~mm$^2$s$^{-1}$, while Linseis Aprosoft~($a=39.1$~mm$^2$s$^{-1}$) lacks that capability; \item an experiment at~$T_0 = 973$~K. Although the curves look similar, the model calculations in PULsE~($a=41.1$~mm$^2$s$^{-1}$) are in closer agreement with the measured data compared to Linseis Aprosoft~($a=43.2 \ \mathrm{mm}^2 s^{-1}$) \end{enumerate*}.}
	\end{figure*}
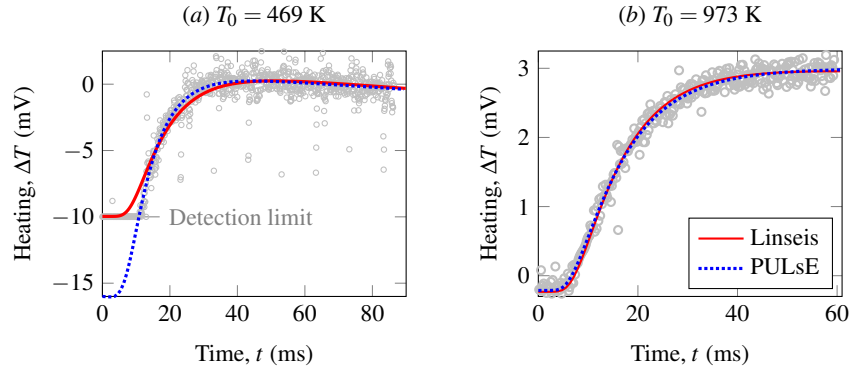

	Experimental data was processed individually for each curve, adjusting the time range only when such intervention was necessary; an example of this is shown in Fig.~\ref{fig:validation_example} (\textit{a}), and a standard truncation routine was applied automatically~(see Sect.~\ref{sect:domain}). 
	
\begin{figure*}
		\centering
		\begin{tikzpicture}
		\begin{groupplot}
		[group style={
			group name=convergence,
			group size=2 by 1,
			xlabels at=edge bottom,
			ylabels at=edge left,
			horizontal sep=60pt,
		},
		width=0.65\columnwidth,
		ylabel near ticks,
		]
		\nextgroupplot[ylabel={Coef. of determination, $R^2$},xlabel={Temperature, $T_0$ (K)},title={(\textit{a})},xtick distance = 600]
		\addplot+[ultra thick,color=blue,mark=square,only marks,
		error bars/.cd,y dir=both,y explicit] 
		table[x expr=\thisrowno{2},y expr=\thisrowno{10}, y error expr=\thisrowno{11}] {AverageResults_3.csv};
		\nextgroupplot[ymin=0,ybar,legend pos=north west,ylabel={Probability density},xlabel={Error, $a - a_{\mathrm{ref}}$ (mm$^2$s$^{-1}$)},title={(\textit{b})}]
		\addplot +[
		hist={
			density,
		} 
		] table [y index=0] {Deviation_Comparison};
		\addlegendentry{Linseis};
		\addplot +[
		hist={
			density,		
		},   
		opacity=0.725  
		] table [y index=1] {Deviation_Comparison};
		\addlegendentry{PULsE};
		\end{groupplot}		
		\end{tikzpicture}
		\caption[foo_bar]{\label{fig:validation_analysis}Statistical analysis of the thermal diffusivity data: \begin{enumerate*}[label=(\alph*)] 
				\item the coefficient of determination~$R^2$ calculated with PULsE and averaged at each test temperature, showing a correlation with the detector signal-to-noise ratio;
				\item the probability density of error~$a_i - a_{i,\mathrm{ref}}$, where~$a_{i,\mathrm{ref}}$ are the Netzsch reference values, with the median values~$0.867$ and~$-0.18$~mm$^2$s$^{-1}$ for thermal diffusivity data obtained with the Linseis Aprosoft and PULsE software respectively
			\end{enumerate*}.}
	\end{figure*}
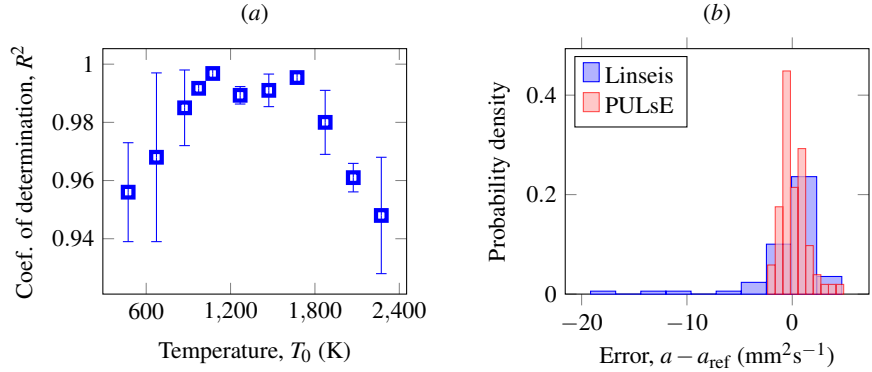

	The default search variables were: the thermal diffusivity~$a$, the maximum heating~$\Delta U_{\mathrm{max}}$, and the baseline intercept~$T_{\mathrm{lin}}$. A linear negative drift of the detector signal could be observed at~$T_0 = 473$~K~(see e.g.~Fig.~\ref{fig:validation_example} (\textit{a})), which required including the baseline slope as a search variable at that temperature. This drift likely originates from the heat exchange between the sample and the sample holder~\cite{philipp2019accuracy}~(either by conduction, radiation, or both). If included in the search vector, the latter would have been poorly estimated due to a low signal-to-noise ratio~(see Fig.~\ref{fig:validation_analysis} (\textit{a})). At~$T_0 \leq 1073$~K and~$T_0 > 1873$~K the heat losses were indistinguishable from the detector noise~(e.g.~Fig.~\ref{fig:validation_example}~(\textit{b})). Likewise, at medium temperatures~($1273 \leq T_0 \leq 1873$~K) the accuracy of detector measurements was sufficiently high and the temperature-dependent heat losses pronounced. All of these factors had to be taken into account manually when performing the final calculations~(see supplementary material). Unfortunately, PULsE still requires manual input based on the recommendations above to produce the most accurate results. Work is currently undergoing to deliver a fully-automatic procedure for determining the most important independent variables based on a statistical data analysis.
	
	Finally, a residual analysis was conducted for the thermal diffusivity datasets from the Linseis Aprosoft and PULsE software~(shown in Fig.~\ref{fig:validation_analysis} (\textit{b})). A large number of outliers characteristic to the Aprosoft results~(Fig.~\ref{fig:validation_summary}) was due to an intermittent detector failure, with the signal saturating at either the lower~($-10$~mV) or higher~($+10$~mV) detection limits~(this could have also been due to an operator error when selecting the detector gain). PULsE, on the other hand, was able to reconstruct the heating curve based on these incomplete measurements by limiting the search range, hence the error distribution is localized near zero for PULsE data~(Fig.~\ref{fig:validation_analysis}). Quite importantly, even though the commercial software was able to deliver meaningful values in many cases, the associated error distribution had a median at~$0.867$~mm$^2$s$^{-1}$, meaning that an uncompensated systematic error was present. The median error for the PULsE data was significantly lower~($-0.18$~mm$^2$s$^{-1}$), suggesting a better quality of the search procedure, a better thermal transfer model, or both.
	
	\section{Current limitations of the computational method}
	
	The following list of problems will be addressed in future publications.
	\begin{itemize}
		\item A source of uncertainty associated with non-uniform heating has not been considered in the present study, but is listed in the ASTM document~\cite{astm2013standard}. If the diameter of the laser spot~$d_{\mathrm{las}}$ is smaller than the diameter of the sample~$d$, radial heat fluxes will induce a change to the heating curve~(e.g. as shown by~\citet{Baba_2001}). Hence, a fully two-dimensional heat conduction problem should be used instead;
		\item An automated procedure for selecting an optimal number of variables to achieve the least influence of the systematic experimental  errors on the values of thermal properties is currently lacking; instead, some parameters are selected manually. A better procedure will likely involve a constrained version of nonlinear optimization~(to limit~$\mathrm{Bi} \geq 0$) and a variable time domain. The latter will need to rely on a different statistic than the~$\chi^2$ -- a possible solution would be using a Bayeseian information criterion, which has proven to be effective for a different problem considered recently by~\citet{FULTON2019108221}; 
		\item \citet{Parker1961} have originally estimated that in the adiabatic case the maximum front surface temperature of the sample after a laser pulse is~$T_{\rm{{}max}}(x=0) = 38 T_{\mathrm{max}}(x=l) \cdot l / a^{1/2}$, where $[l] = \textrm{cm}$, $[a] = \rm{cm}^2 \rm{s}^{-1}$. When heat dissipates by radiative transfer, the front-surface heating will be lower; however, it is clear that for sufficiently low test temperatures~$\Delta T/T_0$ may no longer be considered small. The other side of the problem consists in the intrinsic nonlinear dependence of heating on the spectral radiance, which is especially pronounced when~$\Delta T > 30-40$~K~(\citet{Wang2000}). This means that both the nonlinear detector output and the nonlinear heat losses may need to be taken into account for some cases, e.g. for the setup used by~\citet{PAVLOV201739} and~\citet{Ronchi1999};
		\item Current analysis does not cover: multi-layered heat transfer, a distributed heat source, and possible modifications to the original temperature detection concept.  
	\end{itemize}
	
	It is hoped that some of the limitations may be resolved in collaboration with other researchers.
	
	\section{Conclusions}
	
	The following typical errors have been identified by analysing a large array of experimental data from two different instruments:
	
	\begin{enumerate}[label=(\alph*)]
		\item Partial interruption of data acquisition from the detector, e.g. due to an electromagnetic fault or because of the wrong gain setting, resulting in an incomplete measurement;
		\item Synchronization failure causing a time shift in the raw data;
		\item Periodically-occurring outliers, which are suspected to originate from the imperfect soldering used for electric connections;
		\item Superimposed noise harmonics caused by a combination of factors including e.g. laser reflections due to a cracked sample holder cap, mains hum, mechanical vibrations, etc.;
		\item High-amplitude white noise caused by low specific detectivity of the detector or a poorly transmitting optical window;
		\item Unstable data acquisition~(e.g. due to overheating of electric connections to the main amplifier) causing a linear baseline drift;
		\item Conductive or radiative heat transfer between the sample and the sample holder.
	\end{enumerate}
		
	A computational method based on nonlinear optimization and finite-difference schemes~(verified against some standard analytical solutions) has been designed specifically to handle data processing in experiments showing the above-listed problems. The implementation of this method in the PULsE software allows to control the search vector for data processing~(by including or excluding search variables) and to target specific parts of the raw data -- features not commonly present in the commercial software. A built-in tool for the residual analysis helps to identify systematic errors in the processed data.
	
	A cross-validation study has been conducted with data generated using Linseis equipment at the Materials Research Facility~(UKAEA) on a Netzsch reference sample by benchmarking against the packaged Linseis software. Generally, PULsE outperforms Linseis Aprosoft due to wider data treatment capabilities and fine tuning. Future work will focus on extending its capabilities and improving batch-processing of data.    	

	\begin{acknowledgements}
	
	\noindent This work was partially funded by the RCUK Energy Programme~(Grant No. EP/T012250/1). The research used UKAEA's Materials Research Facility, which has been funded by EPSRC and is part of the UK's National Nuclear User Facility and Henry Royce Institute for Advanced Materials. 
	
	\end{acknowledgements}
	
	\section*{Supplementary Material}
	
	\noindent See supplementary material for a complete set of experimental data and calculation results using PULsE and Linseis Aprosoft.
	
	
	\section*{Data Availability Statement}	

\noindent	The data that supports the findings of this study are available within the article and its supplementary material. The software incorporating the algorithms described in this work is openly available at \hyperref[doi:10.5281/zenodo.3708077]{https://doi.org/10.5281/zenodo.3708077}. 
	
	\bibliography{Bibliography}
	
\end{document}